\documentclass[twocolumn, aps,prx,showpacs,bibfootnote,preprintnumbers, mathtools, superscriptaddress, amsmath,amssymb]{revtex4-2}

\usepackage{amsmath,amsfonts,amsthm,bm}
\usepackage{braket}
\usepackage[caption=false]{subfig}
\usepackage{graphicx}
\usepackage{xcolor}
\usepackage{physics}
\usepackage{tikz}
\usetikzlibrary{tikzmark}
\usepackage{dashbox}%
\usepackage{mathtools, amssymb}
\usepackage[colorlinks=true]{hyperref}
\usepackage{orcidlink}
\usepackage{changes}

\newcounter{cycle}

\begin{document}
\title{Quantum dynamical emulation of imaginary time evolution}

\begin{abstract}
We introduce a constructive method for mapping non-unitary dynamics to a weighted set of unitary operations. We utilize this construction to derive a new correspondence between real and imaginary time, which we term Imaginary Time Quantum Dynamical Emulation (ITQDE). This correspondence enables an imaginary time evolution to be constructed from the overlaps of states evolved in opposite directions. We develop ITQDE as a tool for estimating the ground and thermal state properties associated with a given Hamiltonian. We additionally provide a prescription for leveraging ITQDE to estimate the complete Hamiltonian spectrum. We go on to develop a quantum algorithm for computing Hamiltonian spectra based on ITQDE, which we validate through numerical simulations and quantum hardware implementations. We conclude with a discussion of how ITQDE can be utilized more broadly to derive novel thermodynamic results, including a generalisation of the Hubbard-Stratonovich transformation.  
\end{abstract}

\author{Jacob M. Leamer\orcidlink{0000-0002-2125-7636}}
\affiliation{Tulane University, New Orleans, LA 70118, USA}
\affiliation{Quantum Algorithms and Applications Collaboratory, Sandia National Laboratories, Albuquerque, New Mexico 87185, USA}

\author{Alicia B. Magann\orcidlink{0000-0002-1402-3487}}
\affiliation{Quantum Algorithms and Applications Collaboratory, Sandia National Laboratories, Albuquerque, New Mexico 87185, USA}

\author{Denys I. Bondar\orcidlink{0000-0002-3626-4804}}
\affiliation{Tulane University, New Orleans, LA 70118, USA}

\author{Gerard McCaul\orcidlink{0000-0001-7972-456X}}
\affiliation{Department of Physics, Loughborough University, Loughborough, UK}
\email{g.mccaul@lboro.ac.uk}

\date{February 2024}

\maketitle

\section{Introduction} 

From their first use in the solution of cubic polynomials \cite{Needham1998-qa}, the introduction of imaginary numbers dramatically expands the scope of both the solvable and conceivable. In the domain of physics, for example, the Wick rotation \cite{wick} is celebrated as a tangible link between statistical mechanics and quantum dynamics, where oscillation is transformed to decay. In this sense, evolution in imaginary time might rightly be thought of as the dynamics of thermalisation.

The link between imaginary time and equilibrium information is derived from the fact that a quantum state evolved in imaginary time $\tau$ will have its higher energy states decay exponentially faster than lower energy states. As long as the system has some initial support on the ground state, then in the limit of large $\tau$, the system converges to this state. Ground state calculations are central to numerous applications \cite{needs_continuum_2010, lehtovaara_solution_2007, chin_any_2009}, e.g., in solid state physics and quantum chemistry \cite{Tasaki1998, ren_towards_2023, koster_ground_2011, ceperley_ground_1980}. Beyond ground states, imaginary time evolution can also be used to calculate spectral gaps \cite{leamer_spectral_2023} and thermal (Gibbs) states \cite{chen_hybrid_2020, sinha_efficient_2024, kadow_isometric_2023}. Such states are crucial in a wide range of inquiry, ranging from the thermalization of quantum systems \cite{PhysRevLett.106.040401, Gogolin_2016, Cramer_2012, PhysRevB.97.134301, shirai_floquetgibbs_2018} to optimization \cite{doi:10.1126/science.220.4598.671, doi:10.1073/pnas.0703685104}.  

It is expected that in the future, quantum computers will advance our ability to perform simulations of quantum systems \cite{low_opt_2017, jiang2018quantum}, including for the purpose of calculating the aforementioned ground, thermal, and excited states. Imaginary time evolution, however, does not map naturally to implementation on (circuit-model) quantum computers. Consequently, the many simulation techniques predicated on imaginary time evolution cannot be directly imported into a quantum algorithms setting. This is due to the fact that quantum algorithms are conventionally represented as unitary operations, while imaginary time evolution is not a unitary process.  Nevertheless, there has been substantial interest in developing quantum algorithms to simulate imaginary time evolution \cite{post-selection-Booth, turro_imaginary-time_2022, mcardle_quantum_2020, kondappan_imaginary-time_2023, kosugi_exhaustive_2023, jouzdani_alt_2022, hejazi_adiabatic_2023}. Examples include variational quantum algorithms \cite{gomes_adaptive_2021, gomes_efficient_2020, mcardle_variational_2019},  quantum imaginary time evolution (QITE) \cite{motta_determining_2020, kamakari_digital_2022, tsuchimochi_improved_2023, sun_quantum_2021} probabilistic imaginary time evolution (PITE) \cite{PhysRevResearch.4.033121,kosugi_exhaustive_2023, turro_imaginary-time_2022, nishi_optimal_2023, xie_probabilistic_2022} algorithms, and approaches based on the linear combination of unitaries (LCU) framework \cite{an_linear_2023}. 

Here, we propose a new, constructive approach for developing representations of non-unitary operations as a weighted set of unitary transformations, which we call {Quantum Dynamical Emulation} (QDE). We apply QDE to the problem of generating imaginary time-like evolutions and derive a correspondence we term Imaginary Time Quantum Dynamical Emulation (ITQDE). Using ITQDE, we show that expectation values of a state evolved in imaginary time can be expressed in terms of the overlaps of states that are propagated forwards and backwards in real time. The resulting correspondence establishes novel links between unitary evolutions and \textit{Gaussian}, rather than Gibbs, states. This result is generically applicable to a number of distinct contexts, and in the present work we demonstrate that ITQDE can be leveraged to perform sampling-based calculations of Hamiltonian spectra on quantum computers. To this end, we develop ITQDE into a quantum algorithm, and illustrate its implementation in numerical experiments and in superconducting qubit-based quantum devices. 

The remainder of the paper is organised as follows. We begin by deriving the ITQDE correspondence for representing imaginary time dynamics in terms of unitary operations in Sec.~\ref{sec:derivation}. The analytical and numerical properties of ITQDE are discussed, together with its continuous-time limit.  In Sec.~\ref{sec:stomp}, we develop a technique for using ITQDE to estimate Hamiltonian spectra from dynamical overlaps. Sec.~\ref{sec:numericalillustrations} then provides numerical illustrations of spectral calculations enabled by ITQDE. A first quantum hardware demonstration utilising this correspondence is then presented in Sec.~\ref{sec:quantum computer}, along with a quantum algorithm for implementing ITQDE. We then sketch a number of potential applications for the ITQDE correspondence beyond spectral calculations in Sec.~\ref{sec:Thermal}. We close the paper with a discussion of future directions and generalisations of this work in Sec.~\ref{sec:outlook}.

\section{Imaginary time quantum dynamical emulation \label{sec:derivation}}

In this section, we derive the ITQDE correspondence that allows for emulating imaginary time dynamics using weighted sets of unitary operations. The foundation of the correspondence is the definition of the superoperator
\begin{equation}
\label{eq:superoperator}
    \mathcal{L}[\rho] = \sum_{ij} k_{ij} \mathcal{U}_i\rho\mathcal{U}_j,
\end{equation}
where $k_{ij}$ are scalar coefficients and $\{\mathcal{U}_j\}$ denotes a set of unitary operations. In the case that $\mathcal{U}_j = \delta_{ij}\mathcal{U}_i^\dagger$, if we further stipulate that $\sum_{i} k_{ii} \mathcal{U}_i\mathcal{U}^\dagger_i = 1$, we observe that $\mathcal{L}$ has the form of a set of Kraus operators ~\cite{1983} and constitutes a completely-positive trace-preserving (CPTP) map ~\cite{wilde2013quantum}. Note, however, that non-unitary operations (e.g., imaginary time evolution) are not necessarily trace preserving. If we wish to emulate such operations, we must allow for a choice of unitaries for which $ \mathcal{L}[\rho]$ is {not} CPTP,  and therefore does not correspond directly to a physical evolution. In such a case, we emphasise that while $\mathcal{L}$ is not physical, it is constructed as a weighted sum of unitary operations, each of which are individually realizable physically. 

\begin{figure}
    \centering
\includegraphics[width=1\linewidth]{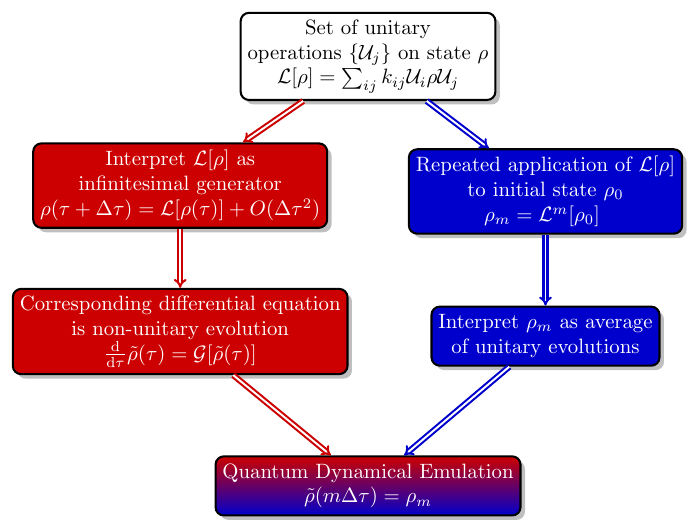}
 \caption{Schematic illustration of Quantum Dynamical Emulation (QDE). The action of a set of unitary operations on a state has an alternative interpretation as a generator for differential equation. Depending on the chosen form of $\{\mathcal{U}_j\}$, this may be a non-unitary evolution. QDE equates the result of such a non-unitary evolution with the unitary operation $\mathcal{L}^m[\rho]$.}
    \label{fig:schematic}
\end{figure}

To demonstrate this premise, we stipulate that each unitary is parametrised by a scalar increment $\Delta \tau$. This allows for the  interpretation of $\mathcal{L}[\rho]$, as an infinitesimal generator for a differential equation. Specifically, $\mathcal{L}[\rho]$ can be Taylor expanded to the form
\begin{equation}
    \mathcal{L}[\rho]=\rho +\Delta \tau \mathcal{G}[\rho] +\mathcal{O}(\Delta \tau^2),
\end{equation}
such that $\mathcal{L}^m[\rho]$ corresponds to a state that is the solution to the differential equation 
\begin{equation}
\frac{{\rm d}}{{\rm d} \tau}\tilde{\rho}(\tau)=\mathcal{G}[\tilde{\rho}(\tau)]   
\end{equation}
for $\tau =m\Delta \tau$. Even in cases where $\mathcal{L}^m[\rho]$ cannot be realized directly, Eq. (\ref{eq:superoperator}) can be interpreted as a weighted average of unitary operations on the state $\rho$, each of which is individually implementable. Specifically, while $\mathcal{L}[\rho]$ may not be directly implementable, the expectation $\rm {Tr}[\mathcal{U}_i\rho \mathcal{U}_j]$ can always be cast as a sum of overlaps between states evolved by the unitaries $\mathcal{U}_i$ and $\mathcal{U}_j$, which are experimentally accessible. This premise is depicted depicted in Fig. \ref{fig:schematic}, where on the level of expectations it is possible to interpret the solution generated by $\mathcal{L}[\rho]$ as an average of unitarily evolved states. This correspondence builds on previous work on ensemble rank truncation~\cite{McCaulensemble} and corner space techniques~\cite{le_bris_low-rank_2013, finazzi_corner-space_2015, donatella_continuous-time_2021}, where it was conceived of as an efficient deterministic method for the simulation of Lindblad equations. QDE extends this idea to the representation of more general forms of dynamics where the CPTP property does not necessarily hold. Having established the motivating arguments for QDE, we now deploy it to derive the ITQDE correspondence. 

To build towards developing the ITQDE representation of imaginary-time dynamics, we begin by considering the unitary transformation 
\begin{equation}
	\mathcal{U} = e^{-i\sqrt{\frac{\Delta\tau}{2}}H},
 \label{eq:U}
\end{equation}
which represents evolution in real time under a Hamiltonian $H$ for a time $\sqrt{\frac{\Delta\tau}{2}}$. With this, we define the superoperator 
\begin{align}
	\mathcal{L}[\rho] &= \frac{1}{2}\left({\mathcal{U}}\rho(\tau){\mathcal{U}} + {\mathcal{U}}^\dag\rho(\tau){\mathcal{U}}^\dag\right),
 \label{eq:UrhoU}
\end{align}
whose effect at first order in $\Delta\tau$ is given by
\begin{align}
\mathcal{L}[\rho]&= \rho - \frac{\Delta\tau}{4}\left(\big\{\rho, H^2\big\} + 2H\rho H\right) + \mathcal{O}(\Delta\tau^2),
\label{eq:firstordersuperoperator}
\end{align}
where $\{\cdot, \cdot\}$ denotes the anticommutator.  From Eq.~\ref{eq:firstordersuperoperator}, we observe that for $\Delta\tau\rightarrow 0$, $\mathcal{L}[\rho]$ corresponds to an infinitesimal evolution of $\rho$ according to 
\begin{align}
	\frac{d\rho}{d\tau} = -\frac{1}{4}\left(\{\rho, H^2\} + 2H\rho H\right), \label{eq:diff}
\end{align} 
whose (unnormalized) solution is given by
\begin{align}
	\rho(\tau) = \sum_{ij}e^{-\frac{1}{4}\tau(E_i+E_j)^2}p_{ij}(0)|E_i\rangle\langle E_j|.
 \label{eq:rhosolution}
\end{align}
In Eq.~\ref{eq:rhosolution}, $\rho(\tau)$ is expressed in the eigenbasis of the Hamiltonian $H = \sum_i E_i|E_i\rangle\langle E_i|$ with general initial condition $\rho(0) = \sum_{ij} p_{ij}(0)|E_i\rangle\langle E_j|$. 

In the limit of large $\tau$, the dominant contribution in $\rho(\tau)$ will be the eigenstate of $H$ with the smallest squared eigenvalue, i.e., 
\begin{equation}
    \rho(\tau) \rightarrow e^{-\tau E_\ell^2}p_{\ell\ell}(0)|E_\ell\rangle\langle E_\ell|
    \label{eq:infinitelimit}
\end{equation}
for $\tau\rightarrow\infty$, where $E_\ell^2 = \min_j E_j^2$ is the ground state energy of $H^2$. Other matrix elements of $\rho(\tau)$ will be exponentially suppressed in $\tau$ according to Eq.~\ref{eq:rhosolution}. 

The result in Eq.~\ref{eq:infinitelimit} is directly analogous to the outcome of evolving the state $\rho(0)$ in \emph{imaginary time} under the Hamiltonian $H^2$, and could accordingly be used to study ground states of $H^2$ via Eq. (\ref{eq:infinitelimit}). However, we emphasize that we have obtained Eq.~\ref{eq:infinitelimit} utilizing purely unitary, real-time evolutions per Eq.~\ref{eq:UrhoU}.

Having obtained Eq.~\eqref{eq:rhosolution}, we now  consider the repeated application of Eq.~\ref{eq:UrhoU}, denoting $m$ applications of $\mathcal{L}$ by $\mathcal{L}^m[\rho]$. We consider an initial condition that is given by the pure state 
\begin{align}
	\rho(0) = |\psi_0\rangle\langle\psi_0|,
\end{align}
noting that the generalization to mixed states is straightforward.  Introducing the following notation for forwards and backwards evolved states,
\begin{align}
	|\psi_{j+1}\rangle &= {\mathcal{U}}|\psi_j\rangle,  & |\psi_{j-1}\rangle &= {\mathcal{U}}^\dag|\psi_j\rangle,
\end{align}
 a single application of $\mathcal{L}$ yields
\begin{equation}
	 \rho(\Delta \tau) = \mathcal{L}^1[\rho(0)] = \frac{1}{2}(|\psi_1\rangle\langle\psi_{-1}| + |\psi_{-1}\rangle\langle\psi_1|).
\end{equation}
The effect of repeated applications of $\mathcal{L}$ can then be ascertained inductively.  Namely, after $m$ steps, $\rho(m\Delta\tau)$ may be written as
\begin{equation}
	\rho(m\Delta\tau) = \mathcal{L}^m[\rho(0)] = \sum_{j=0}^{2m} a^{(m)}_j|\psi_{j-m}\rangle\langle\psi_{m-j}|, \label{eq:1step}
\end{equation}
where the $a^{(m)}_j$ denote coefficients that remain to be calculated. These coefficients have two important properties: first, $a^{(m)}_0 = a^{(m)}_{2m} = 1$, and second, the lack of a static term in $\mathcal{L}$ guarantees $a^{(m)}_j = 0$ for odd $j$.  

Applying $\mathcal{L}$ again to Eq.~\eqref{eq:1step}, we obtain
\begin{align}
	2\mathcal{L}^{m+1}[\rho(0)] = \sum_{j=0}^{j=2m}a^{(m)}_j(|\psi_{j-(m+1)}\rangle\langle\psi_{m+1-j}| \\
 +|\psi_{j+1-m}\rangle\langle\psi_{m-1-j}|) \nonumber.
\end{align}
Rearranging indices, we find
\begin{align}
	2\mathcal{L}^{m+1}[\rho(0)] = |\psi_{-m-1}\rangle\langle\psi_{m+1}|+|\psi_{m+1}\rangle\langle\psi_{-m-1}| \\ + \sum_{j=2}^{j=2m}\left(a_j^{(m)}+a_{j-2}^{(m)}\right)|\psi_{j-(m+1)}\rangle\langle\psi_{m+1-j}| \nonumber.
\end{align}
This yields the following recursion relation
\begin{align}
	2a_j^{(m+1)} = a_j^{(m)} + a_{j-2}^{(m)},
\end{align}
whose solution is given by Pascal's identity
\begin{align}
	{{m}\choose {k}} = {{m-1}\choose {k}} + {{m-1}\choose {k-1}},
\end{align}
yielding
\begin{equation}
    a^{(m)}_j = \begin{cases}
  \frac{1}{2^m}{m \choose \frac{j}{2}}  & j \ \text{even}, \\
  0 & j\  \text{odd}.
\end{cases}
\label{eq:evenandoddaj}
\end{equation}
Using $\tau=m\Delta \tau$, we substitute Eq.~\ref{eq:evenandoddaj} into Eq.~\ref{eq:1step} to finally obtain
\begin{equation}
	\rho(\tau) = \frac{1}{2^m}\sum_{j=0}^{j=m}{{m}\choose {j}}|\psi_{2j-m}\rangle\langle\psi_{m-2j}|.
 \label{eq:correspondence}
\end{equation}
Eq.~\eqref{eq:correspondence} can be used to simulate the evolution of an initial state $\rho(0)$ in imaginary time under $H^2$ using only unitary operations, and thus enables what we term  \textit{Imaginary Time Quantum Dynamical Emulation} (ITQDE).  The error between Eq.~\eqref{eq:correspondence} and the continuous-time limit given previously in Eq.~\eqref{eq:rhosolution} is $\mathcal{O}(m\Delta\tau^2)$, noting that in the sections below, we do not explicitly retain this error term. 

In many cases, it is desirable to additionally enforce that the state is normalized at all (imaginary) times. This can be accomplished by introducing the partition function, which we denote by $Z(\tau) = \text{Tr}\left({\rho}(\tau)\right)$. We denote the normalized state by $\overline{\rho}(\tau)\equiv \frac{{\rho}(\tau)}{Z(\tau)}$. With this normalization, we can express the expectation value of an observable, $O$, under the state $\overline{\rho}(\tau)$ as 
\begin{equation}
\label{Eq:discrete_expectation}
    \langle O\rangle(\tau)=\frac{1}{2^mZ(\tau)}\sum_{j=0}^{j=m}{ m \choose {j}}\langle\psi_{2j-m}|O|\psi_{m-2j}\rangle,
\end{equation}
where 
\begin{equation}
\label{Eq:partitionfunction}
    Z(\tau)=\frac{1}{2^m}\sum_{j=0}^{j=m}{ m\choose {j}}\langle\psi_{2j-m}|\psi_{m-2j}\rangle,
\end{equation}
noting that when substituting Eq.~\eqref{Eq:partitionfunction} into Eq.~\eqref{Eq:discrete_expectation}, the factors of $2^m$ cancel.

Equations~\eqref{Eq:discrete_expectation} and \eqref{Eq:partitionfunction} illustrate that an observable expectation value with respect to the imaginary time evolved state $\overline{\rho}(\tau)$ is expressible as a weighted sum of overlaps between states that have been evolved unitarily in opposite directions in real time. In order to estimate the expectation value of an observable under the ground state, $|E_\ell\rangle$, of $H^2$, one could utilize Eq.~\eqref{Eq:discrete_expectation} carried out to large $\tau$, in line with Eq.~\eqref{eq:infinitelimit}, where $\rho(0)$ can be any state with nonzero support on $|E_\ell\rangle$. 

\subsection*{Large $m$ limit}
In settings where many time steps are needed to achieve convergence (i.e., the large $m$ limit), evaluating  Eqs.~\eqref{Eq:discrete_expectation} and \eqref{Eq:partitionfunction} can be computationally challenging due to the presence of the binomial coefficients ${{m}\choose j}$, whose magnitudes increase very quickly with $m$. As a practical matter, this challenge can be addressed by first shifting the summation index to make the symmetric nature of the sum explicit, i.e., replacing $\sum_{j=0 }^{j=m} { m\choose {j}}$ by $\sum_{j=-\frac{m}{2} }^{j=\frac{m}{2} } { m \choose {\frac{m}{2}}+j}$ in Eqs.~\eqref{Eq:discrete_expectation} and \eqref{Eq:partitionfunction}, and then considering the large $m$ asymptotic for the binomial coefficient
\begin{equation}
\label{eq:binomialapprox}
    {m \choose m/2 + j } \sim \frac{2^m}{\sqrt{m\pi/2}} e^{-2j^2/m}.
\end{equation}
Henceforth, we assume $m$ to be even. This expression is readily derivable from the Stirling approximation \cite{Spencer2014-se}, with an error on the order $\mathcal{O}(j^3 m^{-2})$, although as a practical matter this error becomes negligible even for relatively small values of $m$. Substituting Eq.~\eqref{eq:binomialapprox} into Eqs.~\eqref{Eq:discrete_expectation} and \eqref{Eq:partitionfunction} yields
\begin{equation}
    \begin{aligned}
    \tilde{\langle O\rangle}(\tau)  &= \frac{1}{\tilde{Z}(\tau)} \sum_{j=0}^{j=\frac{m}{2}} \frac{{ e}^{-\frac{2j^2}{m}} }{\sqrt{m\pi/2}}\langle\psi_{2j}|O|\psi_{-2j}\rangle +\text{c.c.} \label{eq:approx_shifted_expt_nolambda}
    \end{aligned}
\end{equation}
and
\begin{equation}
    \begin{aligned}
  \tilde{Z}(\tau) &= \frac{1}{\sqrt{m\pi/2}}\sum_{j=0}^{j=\frac{m}{2}} e^{-\frac{2j^2}{m} } \langle\psi_{2j}|\psi_{-2j}\rangle+ \text{c.c.}, \label{eq:approx_shifted_z_nolambda}
    \end{aligned}
\end{equation}
where we use the tilde to denote that the large $m$ approximation has been used. Equations~\eqref{eq:approx_shifted_expt_nolambda} and \eqref{eq:approx_shifted_z_nolambda} can be used as alternative equations when $m$ becomes large.

\subsection*{Maximally mixed initial condition}

In the case where the initial condition is given by the a state that is proportional to the identity matrix, $\rho(0) \propto {I}$, a number of further results can be inferred from ITQDE. This initial condition  corresponds to the maximally mixed or infinite temperature state. Beginning from this condition means that at all $\tau$, the squared Hamiltonian, $H^2$, and $\rho$ necessarily commute, simplifying Eq.~\eqref{eq:diff} such that we obtain at time $\tau$ the Gaussian state 
\begin{align}
	{\rho}(\tau) = e^{-\tau H^2},
 \label{eq:gaussianstate}
\end{align}
which is equivalent to a thermal (Gibbs) state with an effective Hamiltonian $H^2$ at inverse temperature $\beta = \tau$ \cite{DMM, bondar_efficient_2016}. This illustrates that in this setting, evolving $\rho(0) \propto I$ under Eq.~\eqref{eq:diff} is functionally equivalent to imaginary time propagation under $H^2$. This suggests that it is desirable to consider this initial condition, given its natural relationship to thermal states. 

Furthermore, given $\rho(0) = I$, the final (unnormalized) state in Eq. (\ref{eq:correspondence}) becomes 
\begin{equation}
	\rho(\tau) = \frac{1}{2^m}\sum_{j=0}^{j=m}{{m}\choose {j}}e^{-i(2j-m)\sqrt{2\Delta\tau}H}
 \label{eq:discretemaxmixedsoln}
\end{equation}
where the approximation error between Eq.~\eqref{eq:discretemaxmixedsoln} and the Gaussian state solution given in Eq.~\eqref{eq:gaussianstate} is also $\mathcal{O}(m\Delta\tau^2)$. In this setting, Eq.~\eqref{Eq:discrete_expectation} can be used to estimate the expectation values of observables under thermal states associated with inverse temperature $\beta = \tau$.

\section{ITQDE for spectral calculations \label{sec:stomp}}

We now detail how the ITQDE correspondence can be used to perform spectral calculations. We first observe that when $O=H^2$, evaluating Eq.~\ref{Eq:discrete_expectation} for large $\tau$ would output an estimate of its ground state energy $E_\ell^2$. Similarly, eigenenergies can be estimated with respect to the thermal state described by Eq.~\eqref{eq:gaussianstate}. To calculate these at some desired inverse temperature $\beta$ one need only implement Eq.~\eqref{Eq:discrete_expectation} to $\tau = \beta$ from an initial condition of $\rho(0) = \frac{1}{Z(0)}I$.

It is, however, possible to go further, and exploit ITQDE to perform full spectral calculations that additionally resolve excited states. To demonstrate this, we first note that for a $d$-dimensional system, there are $2^d$ possible Hamiltonians $H_\kappa$ that square to the same $H^2$.  In other words, for all Hamiltonians $H_\kappa$, parameterized by length-$d$ binary strings $\kappa$ with elements $\kappa_j \in \{0, 1\}$ such that
\begin{equation}
	H_\kappa = \sum_{j=0}^{d-1} (-1)^{\kappa_j}E_j|E_j\rangle\langle E_j|,
\end{equation}
we have that the squares of all these $H_\kappa$ are identically given by 
\begin{equation}
	H^2 = \sum_{j=0}^{d-1}E_j^2|E_j\rangle\langle E_j|.
\end{equation}
This reinforces that as $\tau \rightarrow \infty$, $\rho(\tau)$ will select the eigenstate of $H$ with the smallest squared eigenvalue per Eq.~\eqref{eq:infinitelimit}.  We can take advantage of this behavior by introducing a shift $\lambda$ into the Hamiltonian such that
\begin{equation}
	H^{(\lambda)} = H + \lambda I.
\end{equation}
This shift allows one to directly tune which eigenstate has the lowest squared eigenvalue, and accordingly, which eigenstate $\rho^{(\lambda)}(\tau)$ approaches as the system is evolved to higher $\tau$ \cite{hirsch_dynamic_1983}.  Importantly, the effect of this shift on $\mathcal{U}$ is simply to introduce a global phase of $e^{-i\sqrt{\frac{\Delta \tau}{2}}\lambda}$, i.e., such that $\mathcal{U}^{(\lambda)} \equiv e^{-i\sqrt{\frac{\Delta \tau}{2}}H^{(\lambda)}} = e^{-i\sqrt{\frac{\Delta \tau}{2}}\lambda}e^{-i\sqrt{\frac{\Delta \tau}{2}}H}$.  Given this, we may describe the expectation value of an observable $O$ with respect to $\rho^{(\lambda)}(\tau)$ as
\begin{equation}
\begin{aligned}
	\langle O^{(\lambda)}\rangle(\tau)&=\frac{1}{Z^{(\lambda)}(\tau)}\sum_{j=0}^{\frac{m}{2}}\frac{e^{2i\lambda j\sqrt{\frac{\Delta\tau}{2}}}}{2^m}{{m}\choose {\frac{m}{2}+j}}\langle\psi_{2j}|O|\psi_{-2j}\rangle\\
 &\quad + \text{c.c.}, \label{eq:shifted_expt}
 \end{aligned}
\end{equation}
where
\begin{equation}
	Z^{(\lambda)}(\tau) = \frac{1}{2^m}\sum_{j=0}^{\frac{m}{2}}e^{2i\lambda j\sqrt{\frac{\Delta\tau}{2}}}{{m}\choose {\frac{m}{2}+j}}\langle\psi_{2j}|\psi_{-2j}\rangle + \text{c.c.}, \label{eq:shifted_z}
\end{equation}
and we have exploited the fact that each overlap has a conjugate partner to reduce the number of terms in the sums from $m$ to $\frac{m}{2}$.  

Meanwhile, in the large $m$ limit we obtain
\begin{equation}
    \begin{aligned}
    \langle \tilde{O}^{(\lambda)}\rangle(\tau)  &= \frac{1}{\tilde{Z}^{(\lambda)}(\tau)} \sum_{j=0}^{j=\frac{m}{2}} \frac{{ e}^{-\frac{2j^2}{m} + 2i\lambda j\sqrt{\frac{\Delta\tau}{2}}} }{\sqrt{m\pi/2}}\langle\psi_{2j}|O|\psi_{-2j}\rangle\\
    &\quad +\text{c.c.} \label{eq:approx_shifted_expt}
    \end{aligned}
\end{equation}
and
\begin{equation}
    \begin{aligned}
  \tilde{Z}^{(\lambda)}(\tau) &= \frac{1}{\sqrt{m\pi/2}}\sum_{j=0}^{j=\frac{m}{2}} e^{-\frac{2j^2}{m} + 2i\lambda j\sqrt{\frac{\Delta\tau}{2}}} \langle\psi_{2j}|\psi_{-2j}\rangle+ \text{c.c.}, \label{eq:approx_shifted_z}
    \end{aligned}
\end{equation}
by substituting Eq.~\eqref{eq:binomialapprox} into Eqs.~\eqref{eq:shifted_expt} and \eqref{eq:shifted_z}.

Crucially, when $O = H$, varying the value of $\lambda$ allows us to calculate the entire energy spectrum by simply estimating the overlaps $\langle\psi_{2j}|H|\psi_{-2j}\rangle$ and $\langle\psi_{2j}|\psi_{-2j}\rangle$, $j = 0,\cdots,m/2$ corresponding to a single trajectory in $\tau$, and then evaluating Eq.~\eqref{eq:shifted_expt} for different values of $\lambda$ as a numerically simple post-processing step. We also note that $\lambda$ can be used to accelerate convergence in cases where the ground state is sought.  More details regarding this use are located in App.~\ref{app:accel}.

The ability of ITQDE to resolve spectra is a direct consequence of the imaginary time evolution it implements generating a Gaussian, rather than  thermal (or Gibbs) state. The use of $H^2$ to enable spectral sweeping has been a long-standing proposal in the context of quantum Monte-Carlo methods \cite{PhysRevB.28.5353}. Challenges in this context stem from the fact that $H^2$ is more singular than $H$ for realistic Hamiltonians, and cannot guarantee non-negative weights on paths \cite{Ceperley1988}. The fact that ITQDE never directly employs $H^2$ in its calculation of spectra circumvents this problem, and suggests that its methodology might be usefully adapted and applied to Monte-Carlo calculations.

Lastly, in the case that one seeks to obtain the thermal state directly via ITQDE, it is necessary to obtain the square root of the Hamiltonian, $\sqrt{H}$. While we do not address this in the present work, such an operator will always exist for positive semi-definite $H$, which can itself be guaranteed with a suitable shift of the spectrum.

\section{Numerical illustrations}\label{sec:numericalillustrations}
\begin{figure}
    \centering
    \includegraphics[width=\linewidth]{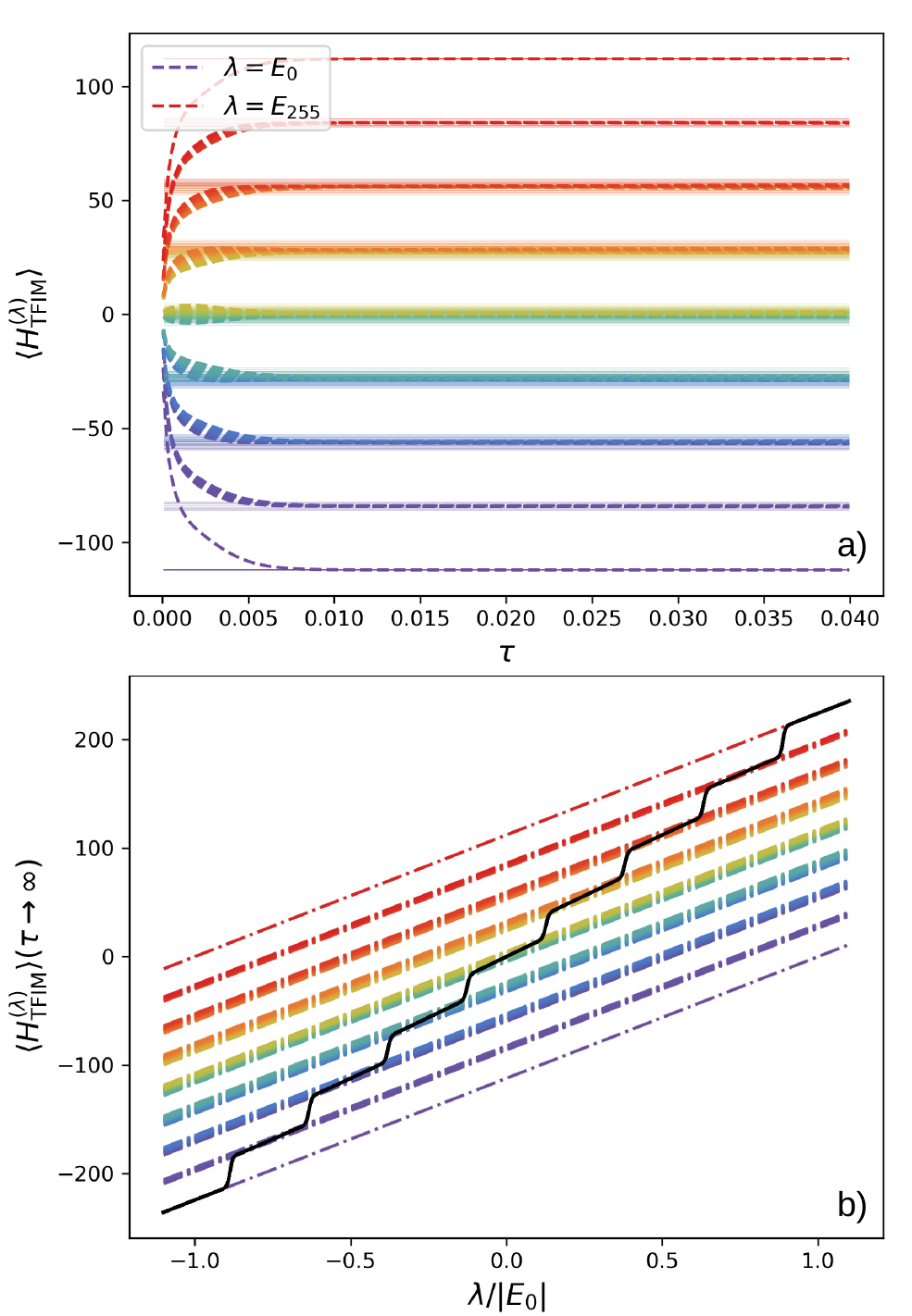}
    \caption{Results of spectral calculations performed for an eight site Ising Hamiltonian using $\Delta\tau=0.4\times 10^{-5}$ and $m=1000$. Dashed lines indicate the eigenenergies of $H_{\mathrm{TFIM}}$. Solid lines indicate the convergence of the spectral calculation to these eigenenergies. For a), we plot the calculated energy as $\lambda$ against propagation in imaginary time $\tau$ and find that our method converges to successively lower eigenstates of $H_{\mathrm{TFIM}}$ before reaching the ground state $E_0$. For b), we plot the change in the steady state expectation value of $H_{\mathrm{TFIM}}$ as $\lambda$ is varied, illustrating that a complete course-grained spectrum is obtainable using only expectation values estimated from a single trajectory only. }
    \label{fig:lambdasweep}
\end{figure}
To demonstrate the capability of Eq.~\eqref{eq:shifted_expt} to resolve the spectrum of $H$, we now illustrate an application of this procedure to the 1D transverse field Ising model, which is a simple model composed of nearest-neighbor coupled spins on a lattice in the presence of an external, uniform magnetic field. It has been studied extensively over the years in the contexts of quantum phase transitions \cite{dziarmaga_dynamics_2005, sachdev_quantum_1999}, quantum spin glasses \cite{kopec_instabilities_1989, laumann_cavity_2008}, and the quantum annealing process \cite{farhi_quantum_2000,troels_2014,shin_how_2014,boixo_evidence_2014} among others.  It is also a useful benchmark problem for developing and testing quantum algorithms, as the model is exactly solvable \cite{weinberg_quspin_2019, dziarmaga_dynamics_2005}, and spins map naturally to qubits. The Hamiltonian is given by
\begin{align}
    H_{\mathrm{TFIM}} = -J\sum_{l=0}^{L-1}Z_{l}Z_{l+1} - h\sum_{l=0}^{L-1}X_l, \label{eq:tfim}
\end{align}
where $Z_l$ and $X_l$ denote Pauli-$Z$ and Pauli-$X$ matrices, respectively, that act on the spin occupying lattice site $l$, $J$ characterizes the strength of the nearest-neighbor interactions, $h$ is the magnetic field strength, and $L$ is the total number of lattice sites. Here, we consider a model with $J=1$, $h=14$, $L=8$, and open boundary conditions. We use Eqs.~\eqref{eq:shifted_expt} and \eqref{eq:shifted_z} to calculate the expectation value $\langle H_{\mathrm{TFIM}}^{(\lambda)}\rangle(\tau)$ for different values of $\lambda$, which are plotted as a function of $\tau$ in Fig.~\ref{fig:lambdasweep}(a).  These results show that changing the value of $\lambda$ causes $\langle H_{\mathrm{TFIM}}^{(\lambda)}\rangle(\tau)$ to converge to different energies, corresponding in this case to the energies associated with each of the different bands of eigenstates.  This premise is further reinforced in Fig.~\ref{fig:lambdasweep}(b), where we plot the steady state expectation value $\langle H_{\mathrm{TFIM}}^{(\lambda)}\rangle(\tau\rightarrow\infty) = \langle  H_{\mathrm{TFIM}}\rangle(\tau\rightarrow\infty) + \lambda$ against $\lambda$, from which the complete spectrum is inferred.
\begin{figure}
    \centering
    \includegraphics[width=\linewidth]{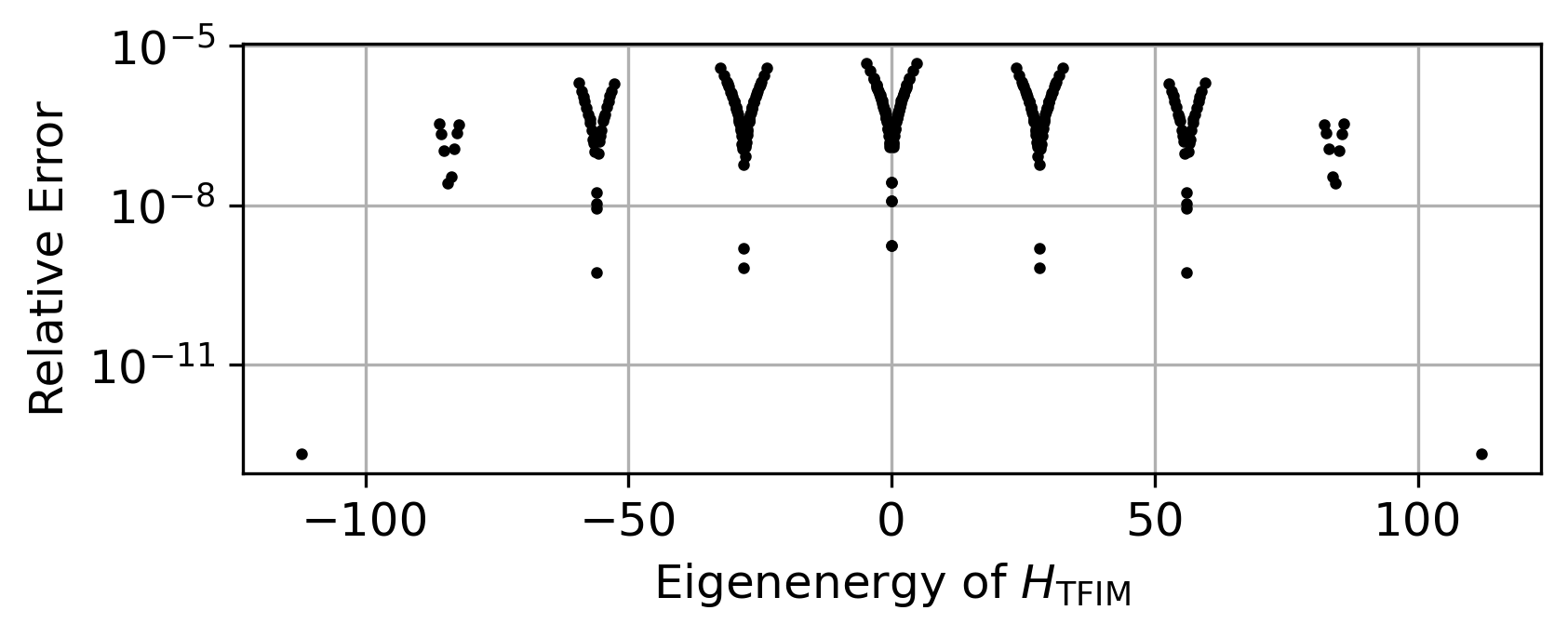}
    \caption{Relative error calculated according to Eq.~\eqref{eq:relerr} in spectral calculations performed using the large $m$ approximation in Eqs.~\eqref{eq:approx_shifted_expt} and \eqref{eq:approx_shifted_z} compared to the results obtained using Eqs.~\eqref{eq:shifted_expt} and \eqref{eq:shifted_z} for the 8 site 1D transverse field Ising model. In both cases, $\Delta\tau=0.4\times 10^{-5}$ and $m=1000$}
    \label{fig:mapprox_err}
\end{figure}

\begin{figure}
    \centering
    \includegraphics[width=\linewidth]{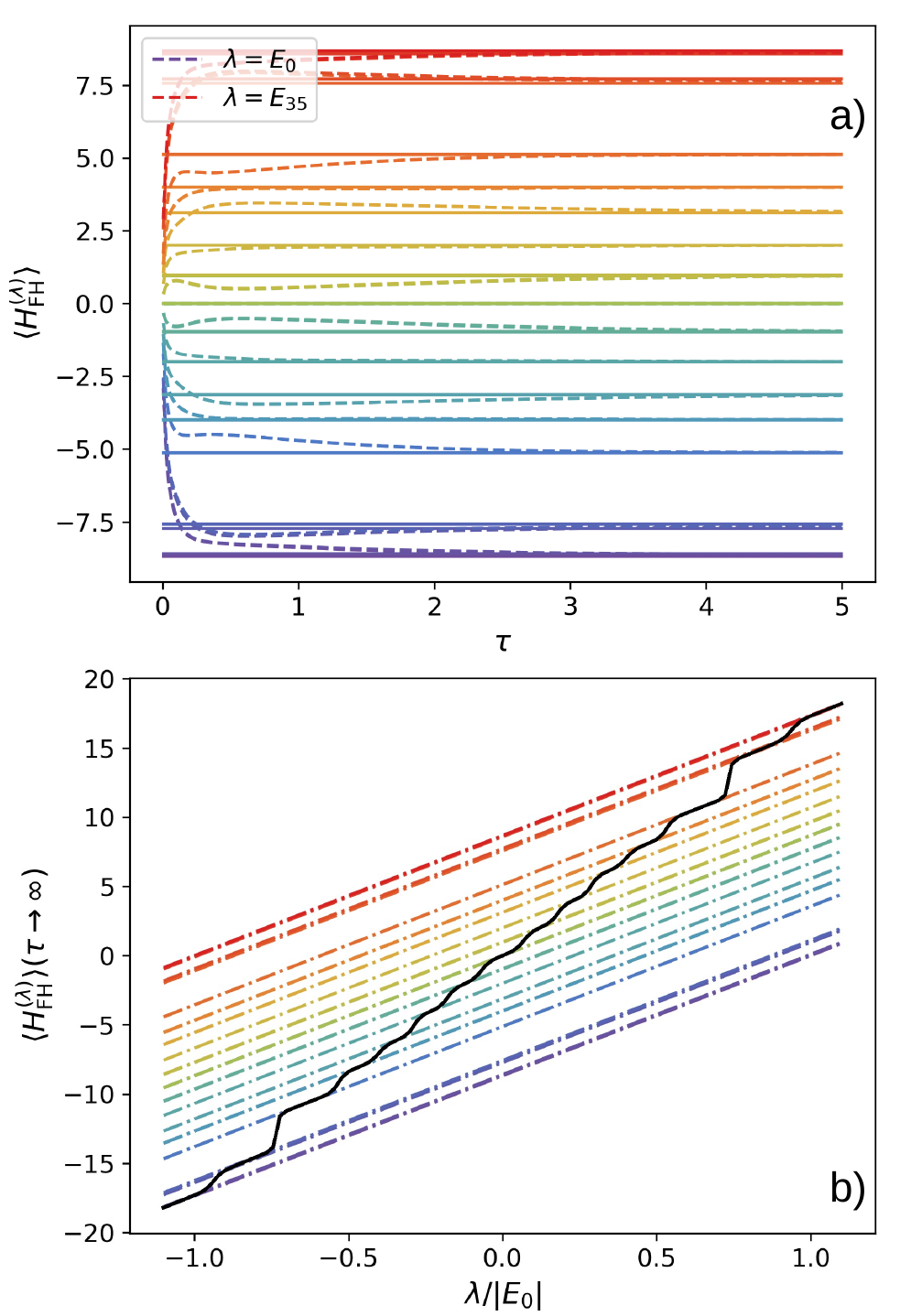}
    \caption{Results of spectral calculations performed for a four site, 2D Fermi-Hubbard model at half-filling with $d\tau = 0.003$ and $m=1500$.  Dashed lines indicate the eigenenergies of $H_{\mathrm{FH}}$, and solid lines show the convergence of $\langle H_{\mathrm{FH}}^{(\lambda)}\rangle (\tau)$ to these eigenenergies.  For a), we plot the calculated energy for various $\lambda$ against $\tau$, and observe good convergence to successively lower eigenstates of $H_{\mathrm{FH}}$ before reaching the ground state, $E_0$.  For b), we plot the change in the steady state expectation value, i.e., $\langle H_{\mathrm{FH}}^{(\lambda)}\rangle (\tau\rightarrow\infty)$ as $\lambda$ is varied, illustrating that a coarse-grained spectrum can be obtained.}
    \label{fig:FH_res}
\end{figure}

\begin{figure}
    \centering
    \includegraphics[width=\linewidth]{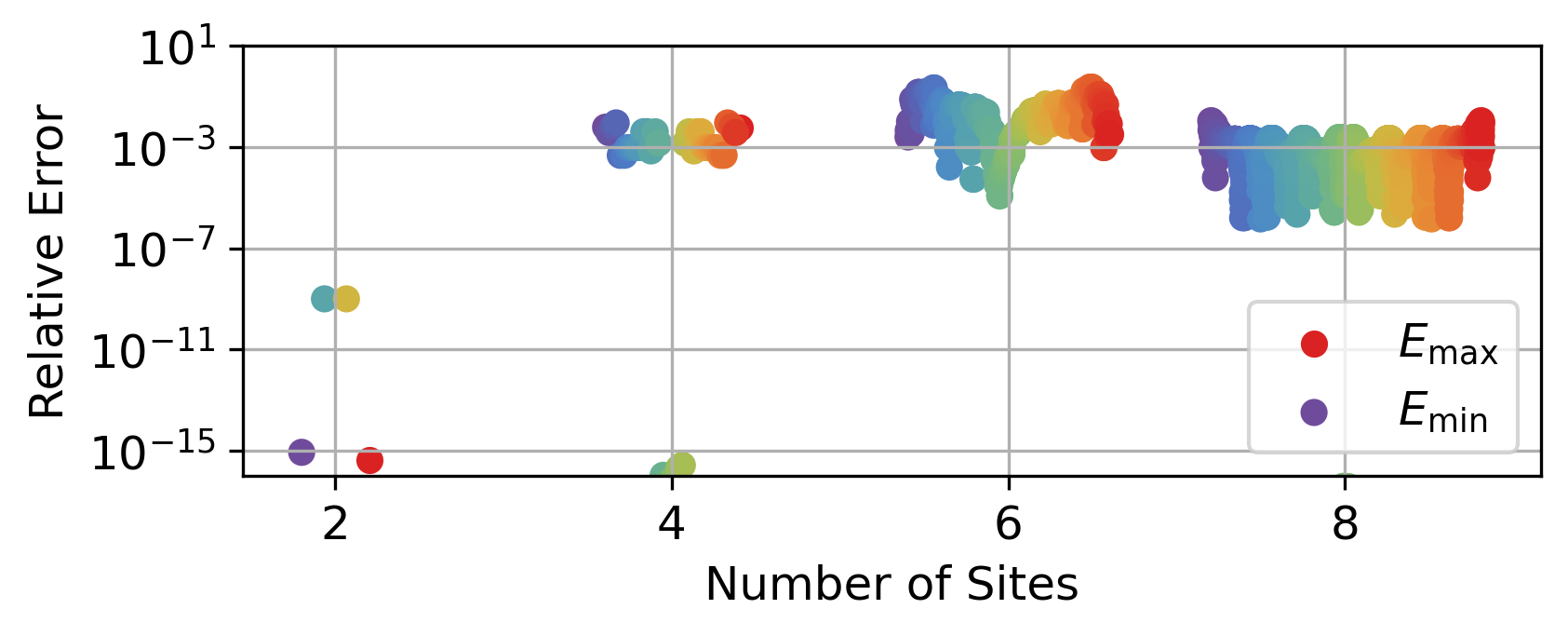}
    \caption{Relative error in spectral calculations of the steady state expectation values for the eigenenergies of the 2D Fermi-Hubbard model at half-filling for 2, 4, 6, and 8 sites. Each eigenenergy error is plotted symmetrically about the spectral centre, which is aligned on the figure with the number of sites that spectrum corresponds to. All calculations were performed using a step size of $\Delta\tau=0.003$ for $m=1500$ steps to a final imaginary time $\tau=5$ with the initial condition $\rho_{\mathrm{FD}}(0) = I$.}
    \label{fig:FH_rel_err}
\end{figure}

We demonstrate the use of the large $m$ approximation by comparing the results obtained using Eqs.~\eqref{eq:approx_shifted_expt} and \eqref{eq:approx_shifted_z} to the ones obtained using Eqs.~\eqref{eq:shifted_expt} and \eqref{eq:shifted_z} for this 8 site transverse field Ising model.  The relative difference
\begin{align}
    \zeta = \frac{|\langle O^{(\lambda)}\rangle - \tilde{\langle O^{(\lambda)}\rangle}|}{|\langle O_g\rangle|}, \label{eq:relerr}
\end{align}
for each eigenenergy of $H_{\mathrm{TFIM}}$ is shown in Fig.~\ref{fig:mapprox_err}.  Here we use $\langle O_g\rangle$ as the minimum eigenenergy of $H_{\mathrm{TFIM}}$ to avoid dividing by small numbers when the calculated eigenvalue is near 0.  This demonstrates that the large $m$ approximation obtains results that are similar to the direct approach using the binomial coefficients. 

We also apply the large $m$ approximation to the 2D Fermi-Hubbard model. The Fermi-Hubbard model is a simple lattice model that aims to capture key aspects of strongly correlated fermionic systems, and has been used to study phase transitions \cite{imada_1998}, superconductivity \cite{anderson_1987}, and magnetism \cite{hofrichter_2016}. The Hamiltonian is given by
\begin{equation}
    H_{\mathrm{FH}} = \sum_{\langle i,j\rangle,\sigma}t_{ij}(c_{i\sigma}^\dag c_{j\sigma} + \text{h.c.}) + U\sum_{j=0}^{L-1} n_{j\uparrow}n_{j\downarrow} - \mu \sum_{j=0}^{L-1} n_{j\sigma},
\end{equation}
where $t_{ij}$ denotes the tunneling amplitude between lattice sites $i$ and $j$, $c_{j\sigma}^\dag$ and  $c_{j\sigma}$ are the creation and annihilation operators associated with a fermion of spin $\sigma$ at lattice site $j$, respectively, $U$ characterizes the on-site interaction of fermions, $n_{j\sigma}=c^\dag_{j\sigma}c_{j\sigma}$ is the occupation number operator, $\mu$ is the chemical potential, and $L$ is the number of lattice sites.  

In Fig.~\ref{fig:FH_res}(a), we plot results showing how the expectation value $\langle H_{\mathrm{FH}}^{(\lambda)}\rangle (\tau)$ varies with $\tau$ and $\lambda$ when the model parameters are set according to $t_{ij}=t=-1$ for all nearest-neighbor pairs, $U=2$, $\mu=0.5$, $L=4$ (corresponding to a $2\times 2$ lattice), and periodic boundary conditions.  These results again demonstrate that changing $\lambda$ causes $\langle H_{\mathrm{FH}}^{(\lambda)}\rangle (\tau)$ to converge to different energies in the spectrum of $H_{\mathrm{FH}}$ and that reasonable results are obtained even when using the large $m$ approximation. The steady state expectation value $\langle H_{\mathrm{FM}}^{(\lambda)}\rangle(\tau \rightarrow \infty)$ of this model for varying $\lambda$ is plotted in Fig.~\ref{fig:FH_res}(b) and further demonstrates that changing $\lambda$ can be used to infer the energetic spectrum.

To examine the performance of our method as the number of sites in the 2D Fermi-Hubbard model increases, we calculate the relative error in the calculation of each eigenenergy compared to the exact values obtained by diagonalizing $H_{\mathrm{FM}}$. This is plotted against the number of sites in the lattice in Fig.~\ref{fig:FH_rel_err}.  We observe from these results that our method is able to accurately calculate the energy eigenvalues of the 2D Fermi-Hubbard model up to 8 sites and that the accuracy is not dependent on the number of sites in the lattice.  The exception is the 2 site case, where much more accurate results are obtained.  This is most likely due to the fact that in the case of 2 sites the model is effectively its much simpler one dimensional equivalent.

All of the results discussed in this section were obtained using the initial condition $\rho(0) = I$.  In App.~\ref{app:diff_ic}, we discuss the effects of choosing a single pure state as the initial condition on the spectral calculation results.

\section{Quantum algorithm for spectral calculations using ITQDE \label{sec:quantum computer}}

\begin{figure}
    \centering
    \includegraphics[width=1.001\linewidth]{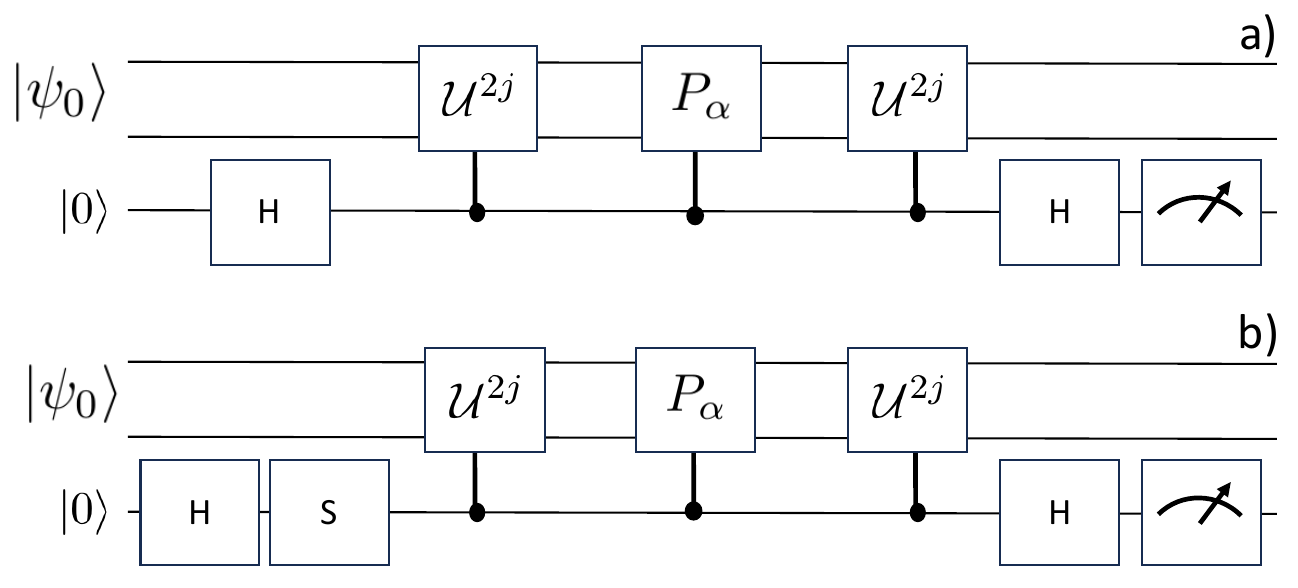}
    \caption{Quantum circuits for performing Hadamard tests to estimate a) $\mathrm{Re}(\langle\psi_{-2j}|P_\alpha|\psi_{2j}\rangle)$ and b) $\mathrm{Im}(\langle\psi_{-2j}|P_\alpha|\psi_{2j}\rangle)$ for use in Eqs.~\eqref{eq:approx_shifted_expt} and \eqref{eq:approx_shifted_z}.   We denote by ${P}_\alpha$ the $\alpha$-th Pauli string in the decomposition of the observable $O$, per Eq. (\ref{eq:overlapwithP}). $H = \frac{1}{\sqrt{2}}\bigl( \begin{smallmatrix}1 & 1\\ 1 & -1\end{smallmatrix}\bigr)$ and $S=\bigl( \begin{smallmatrix}1 & 0\\ 0 & i\end{smallmatrix}\bigr)$ denote the Hadamard and phase gate, respectively.}
    \label{fig:circuit}
\end{figure}

There is currently significant interest in quantum algorithms for simulating imaginary time evolution, with a variety of candidate algorithms developed in recent years. For example, variational quantum algorithms have been developed that aim to optimize a parameterized quantum circuit to best approximate an imaginary time evolution \cite{gomes_adaptive_2021, gomes_efficient_2020, mcardle_variational_2019}. These variational approaches can result in shallow quantum circuits that are favorable for near-term hardware implementations, however, the computational costs of the classical optimization can become intractable for larger systems due to the existence of local minima and barren plateaus. Another leading example is the QITE algorithm \cite{motta_determining_2020, kamakari_digital_2022, tsuchimochi_improved_2023, sun_quantum_2021}, where an evolution in imaginary time is approximated using a Trotter decomposition of a real-time evolution, i.e., where the latter is expressed as the product of unitary, real-time evolutions over a sequence of small time steps. The generators of these real-time evolutions are determined sequentially, based on tomographic measurements performed at each step. This measurement cost can become prohibitive for systems that develop long-range correlations. A third approach is the PITE algorithm \cite{PhysRevResearch.4.033121,kosugi_exhaustive_2023, turro_imaginary-time_2022, nishi_optimal_2023, xie_probabilistic_2022}, where the imaginary time evolution is encoded into a unitary operation acting on a larger Hilbert space. After evolving the total system one step using this larger unitary operator, an ancilla qubit can be measured to determine whether the step of imaginary time evolution was successfully applied or not. If so, the procedure can be repeated to evolve over the next time step, and if not, it must be restarted.  Thus, evolving to later times results in an exponential decrease in the probability of successfully evolving the original system in imaginary time. Some recent works \cite{liu_probabilistic_2021, nishi_acceleration_2022} address this situation by applying amplitude amplification after each step, but the associated costs can become impractical as the number of steps increases. Another recent approach is based on the LCU framework, where the imaginary time evolution is expressed as a linear combination of unitaries \cite{an_linear_2023}.  This approach is conceptually similar to ITQDE, but implements the linear combination of unitaries coherently through the use of ancilla qubits, and the associated overhead costs suggest that it may be most suitable for future fault-tolerant quantum computers. Beyond imaginary time evolution, quantum algorithms for simulating non-unitary dynamics more broadly have been a subject of interest \cite{HamSim, Schrodingerisation}, as have quantum algorithms for computing Hamiltonian spectra. Examples of the latter include the quantum phase estimation algorithm \cite{parker_quantum_2020, kitaev_classical_2002, nielsen_quantum_2010, svore_faster_2013, obrien_quantum_2019, wiebe_efficient_2016, dobsicek_arbitrary_2007}, the rodeo algorithm \cite{choi_rodeo_2021, cohen_optimizing_2023}, methods based on taking the Fourier transform of a time series of expectation values \cite{somma_quantum_2020,gnatenko_detection_2022, gnatenko_energy_2022}, and variational \cite{xie_variational_2024, jones_variational_2019} and feedback-based \cite{rahman2024feedback} quantum algorithms for finding excited states. 

Here, we introduce a new quantum algorithm for calculating spectra via imaginary time evolution that is based on ITQDE. In this formulation, a quantum computer is utilized to estimate the overlaps (i.e., of the form $\langle\psi_{-2j}|\psi_{2j}\rangle$ and $\langle\psi_{-2j}|O|\psi_{2j}\rangle$) that appear in Eqs.~\eqref{eq:shifted_expt}-\eqref{eq:approx_shifted_z}, with the spectrum then calculated by employing these overlaps in classical post-processing. We focus the remainder of this section on how to construct a quantum algorithm for evaluating Eqs.~\eqref{eq:shifted_expt}-\eqref{eq:approx_shifted_z}. Our construction is independent of the initial state $|\psi_0\rangle$, since in practice, $|\psi_0\rangle$ can be selected in a problem-dependent manner. 

There are a number of methods by which the relevant overlaps might be calculated in a quantum computer (see for example App. \ref{app:estimator}). Here, we consider an implementation using the Hadamard test \cite{cleve_quantum_1998} that is detailed fully in App. \ref{app:had_test}. The Hadamard test functions by initializing one ancilla qubit in the state $|0\rangle$ and initializing the primary system register in a state $|\varphi\rangle$. This is followed by Hadamard gates on the ancilla qubit, interleaved with a unitary operation $\mathcal{W}$ that is applied to the primary system register and controlled on the state of the ancilla qubit, and culminating with a final ancilla qubit measurement, as depicted in Fig.~\ref{fig:circuit}(a). Repeating this procedure many times and averaging the results of the final ancilla measurement then outputs an estimate for the real part of the overlap between $|\varphi\rangle$ and $\mathcal{W}|\varphi\rangle$ in terms of the probability, $p_0$, of observing $|0\rangle$  on the ancilla qubit, such that 
\begin{equation}
\text{Re}\left(\langle\varphi|\mathcal{W}|\varphi\rangle\right) = 2p_0 - 1.
\label{eq:realHtest}
\end{equation}
A modified version of the Hadamard test, i.e., one that incorporates an additional phase gate on the ancilla qubit, can be used to obtain the imaginary part of the overlap in an analogous manner, as depicted in Fig.~\ref{fig:circuit}(b).  

We now outline approaches for using the Hadamard test in our setting to estimate $\langle\psi_{-2j}|\psi_{2j}\rangle$ and $\langle\psi_{-2j}|O|\psi_{2j}\rangle$. We begin with the former, which can be expressed as
\begin{equation}
    \langle\psi_{-2j}|\psi_{2j}\rangle = \langle \psi_0|\mathcal{U}^{4j}|\psi_0\rangle.
    \label{eq:firstoverlapHtest}
\end{equation}
Hadamard tests can be performed to estimate the real and imaginary parts of the right-hand-side of Eq.~\eqref{eq:firstoverlapHtest} by associating $\mathcal{W} = \mathcal{U}^{4j} = e^{-i\sqrt{2j\Delta\tau} H}$ and $|\varphi\rangle = |\psi_0\rangle$ in Eq.~\eqref{eq:realHtest}. 

In order to estimate overlaps of the second form, $\langle\psi_{-2j}|O|\psi_{2j}\rangle$, we first expand $O$ in the Pauli operator basis according to
\begin{equation}
    O = \sum_{\alpha=1}^M c_\alpha P_\alpha,
\end{equation}
where $c_\alpha$ and $P_\alpha$ denote a real coefficient and Pauli basis operator, respectively. We note that when $O$ is taken to be a $k$-local observable, $M = \text{poly}(n)$. Importantly, because the Pauli operators are unitary, this expansion allows us to then estimate $\langle\psi_{-2j}|O|\psi_{2j}\rangle$ by taking the sum of a linear combination of constituent overlaps, each evaluated using separate Hadamard tests, according to
\begin{equation}   \langle\psi_{-2j}|O|\psi_{2j}\rangle = \sum_{\alpha=1}^M c_\alpha  \langle \psi_0|\mathcal{U}^{2j} P_\alpha \mathcal{U}^{2j}|\psi_0\rangle,
\label{eq:overlapwithP}
\end{equation}
where for the $\alpha$-th Hadamard test, we have $\mathcal{W} = \mathcal{U}^{2j} P_\alpha \mathcal{U}^{2j}$, and $M$ Hadamard tests are needed in total to evaluate Eq.~\eqref{eq:overlapwithP}. Using this approach to estimate the real and imaginary parts of all $\frac{m}{2}+1$ overlaps in Eqs.~\eqref{eq:approx_shifted_expt} and \eqref{eq:approx_shifted_z} then requires $(m+2)(M+1)=\mathcal{O}(mM)$ separate circuits, each repeated sufficiently many times to ensure convergence of the estimate of the associated overlap. 

Given this Hadamard test formulation, a method for implementing the (real) time evolution $\mathcal{U}^{4j} = e^{-i\sqrt{2j\Delta\tau} H}$ on the primary system register is then all that is additionally required to complete the specification of the quantum algorithm. It is worth noting that in order to simulate imaginary time evolution by an imaginary time $\tau$ under a Hamiltonian $H^2$, we only need real-time evolution under $H$ by time $\sqrt{2\Delta\tau}$. For typical Hamiltonians, $\mathcal{U}^{4j}$ can be implemented {efficiently} on a quantum computer \cite{lloyd1996universal}, and a variety of different Hamiltonian simulation algorithms have been developed for this purpose that could be used. Examples include Suzuki-Trotter product formulas \cite{doi:10.1063/1.529425}, randomized methods like QDRIFT \cite{PhysRevLett.123.070503}, post-Trotter methods \cite{PhysRevLett.114.090502, PhysRevLett.118.010501, low2019hamiltonian, 10.1145/3313276.3316366}, e.g., based on the LCU construction, and hybridized algorithms that combine features of these methods \cite{childs2019faster, Rajput2022hybridizedmethods,hagan2023composite,chakraborty2024implementing,dizaji2024hamiltoniansimulationzenosubspaces}. 

Relative to other quantum algorithms for imaginary time evolution and spectral calculations, as discussed above, the ITQDE-based approach introduced here benefits from the fact that the full energy spectrum can be computed in classical post-processing based on a single imaginary-time trajectory. Furthermore, the ITQDE-based approach does not depend on the preparation of specific or high-fidelity initial states. In addition, classical optimization, tomography, and probabilistic sampling, which can serve to increase the number of circuit repetitions required in variational quantum algorithms, QITE, and PITE, respectively, are not required here. That being said, a large number of circuit repetitions may still be required in the ITQDE-based algorithm in order to resolve each of the overlaps in Eqs.~\eqref{eq:shifted_expt}-\eqref{eq:approx_shifted_z}, and in the future, it would be interesting to compare the sampling costs of the ITQDE-based algorithm with other candidate strategies. Finally, we observe that relative to LCU-based approaches, the ITQDE-based algorithm trades off the need for extra ancilla qubits and deeper circuits for an increase in the number of circuit repetitions. We anticipate that this tradeoff may make ITQDE compatible with nearer-term quantum devices, a prospect that we explore below. 

\subsection*{Quantum device implementations \label{sec:quantum computer2}}

\begin{figure}
    \centering
    \includegraphics[width=\linewidth]{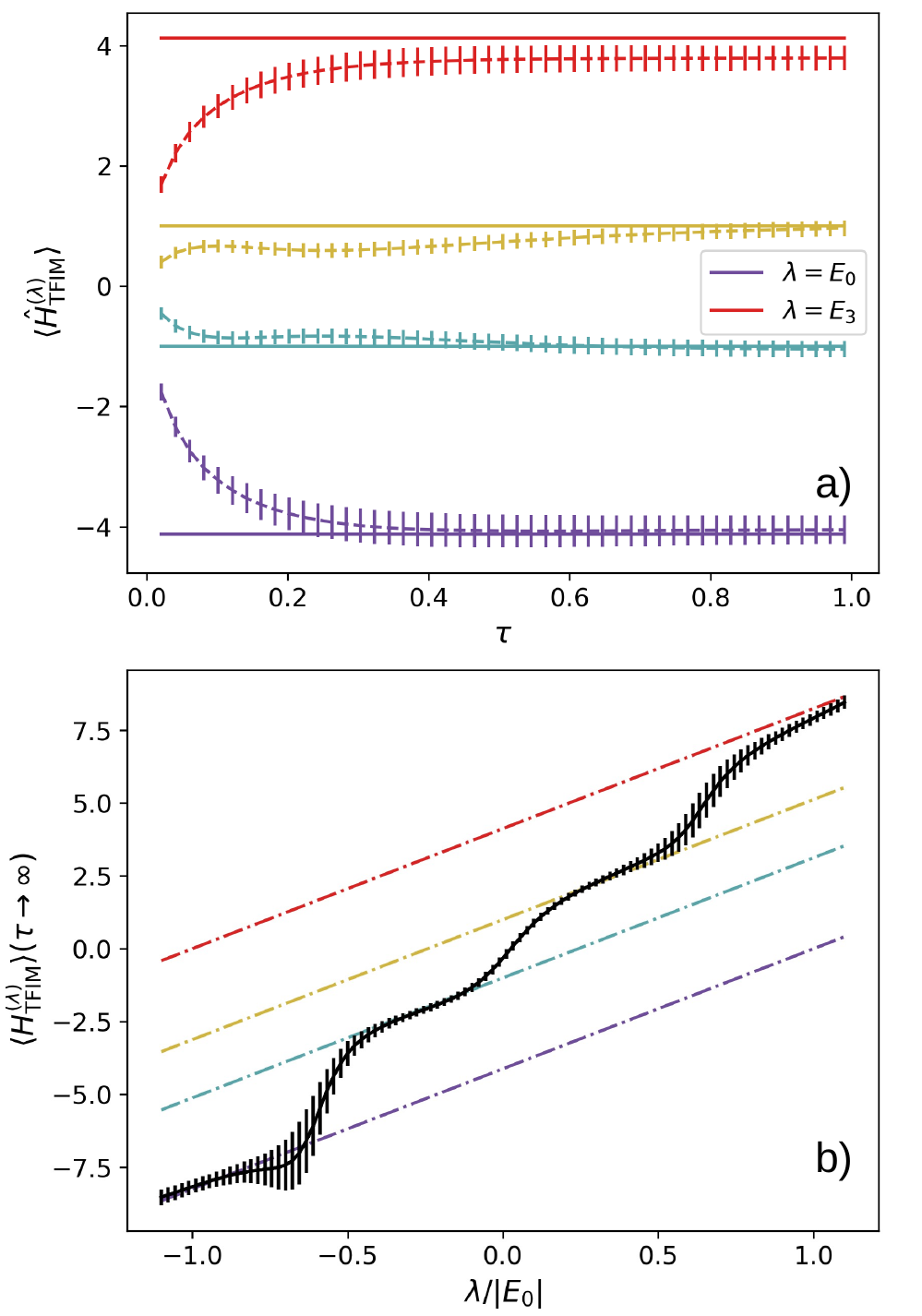}
    \caption{Spectral calculation results for a 2-site transverse field Ising model with $J=1$ and $h=2$, obtained via implementation of ITQDE on an IBM superconducting qubit processor (ibm\_brisbane).  Here we use $\Delta\tau=0.01$ and $m=100$. In a), the expectation values $\langle H_{TFIM}^{(\lambda)}\rangle (\tau)$ output from Eqs.~\eqref{eq:approx_shifted_expt} and \eqref{eq:approx_shifted_z}, with overlaps obtained from ibm\_brisbane, are shown (dashed curves) against the true eigenenergies (solid lines of corresponding color).  Each overlap is obtained using 4000 samples. In b), the change in the steady state expectation of $H_{TFIM}$ as $\lambda$ changes is shown, illustrating that the coarse-grained spectrum can still be inferred when performing ITQDE on quantum hardware.  In both figures, the errorbars denote 95\% confidence.}
    \label{fig:sampler_results}
\end{figure}

Here, we present the results from implementing ITQDE directly on quantum devices. For this, we utilize the IBM superconducting qubit processor ibm\_brisbane, along with IBM's sampler primitive \cite{Qiskit}. Our first implementation considers the transverse field Ising model with $J=1$, $h=2$, $L=2$, and the results are presented in Fig.~\ref{fig:sampler_results}. In this demonstration, the quantum dynamical emulation procedure is repeated for 104 different initial states, prepared by applying random Clifford operations to the initial $|00\rangle$ state, and averaging the overlaps at each step. This averaging procedure (over initial Clifford realizations) aims to produce statistics that converge, with increasing realizations, to those that would be obtained with an initialization in the maximally mixed state.  We expect that the convergence towards the middle energies could be improved by sampling initial states from additional random Cliffords. 
 
\section{Interpretations and Further Applications of ITQDE \label{sec:Thermal}}

Beyond spectral calculations, the character of the ITQDE correspondence has a number of additional properties whose potential implications are discussed in this section. For example, the presence of forward and reverse trajectories in the correspondence is reminiscent of both quantum  \cite{influencefunctional} and stochastic  \cite{Seifert2012}  thermodynamics.  In the latter case, fluctuation theorems are derived from considerations of the probability of trajectories against their reversed counterpart \cite{Spinney2012}. This suggests similar results may be obtained via the methods outlined in previous sections. Indeed, if one were to employ the large $m$ approximation according to Eq.~\eqref{eq:binomialapprox}, the result
\begin{equation}
e^{-\tau {H}^2}=\frac{1}{\sqrt{\pi}}\sum_{j=-\frac{m}{2}}^{j=\frac{m}{2}} \sqrt{\frac{2}{m}}e^{-\frac{2j^2}{m}}e^{-2ij\sqrt{2 \Delta\tau}{H}}
\label{eq:intermediateHSequation}
\end{equation}
can be used to take the expectation of the operator $e^{-\tau H^2}$ with respect to $\ket{\psi_0}$. From this, we directly obtain (reiterating the implicit $\mathcal{O}(m\Delta\tau^2)$ error)

\begin{equation}
\langle e^{-\tau H^2} \rangle = \sqrt{\frac{2}{m\pi}}\sum_{j=0}^{j=\frac{m}{2}} e^{-\frac{2j^2}{m}}\langle\psi_{2j}|\psi_{-2j}\rangle+ \text{c.c.}.
\label{eq:QFT}
\end{equation}
The result in Eq. (\ref{eq:QFT}) bears a similarity to the Crooks fluctuation theorem \cite{PhysRevE.60.2721}, where the work distribution function of a stochastic process can be expressed as a ratio of the probabilities for forward and reversed processes. In Eq. (\ref{eq:QFT}), we find the expected Hamiltonian Gaussian distribution of a state is expressible in terms of an exponentially weighted sum of \textit{overlaps} between forward and backward evolutions from that state. Indeed, the ITQDE derived expression in Eq. (\ref{eq:QFT}) is manifestly quantum, insofar as it utilises overlaps rather than probabilistic ratios, and it links to distributions over $H$, rather than to the non-observable work variable. 

More broadly, it is interesting to observe the behaviour of ITQDE under time reversal. The only sense in which a time increment appears is ${\Delta \tau}$, which in the real-time propagation occurs under a square root. For this reason, the notion of time reversal is most naturally captured by the equivalent transformation of $H \to -H$. When considering the Gaussian distribution, this reversal leaves the ITQDE correspondence unchanged, rendering any question of time directionality redundant. If, however, we consider this transformation in the case of Gibbs states, the necessity of employing  $\sqrt{H}$ in the propagator means this reversal symmetry is broken. That is, direct consideration of thermal states (which are themselves usually understood to be a product of irreversible dynamics) requires the use of a generator that is not invariant under time-reversal. This fact, to the philosophically inclined, may be of some interest.

It is also possible to {invert} the ITQDE correspondence. First, taking the  same large $m$ approximation of  Eq.~\eqref{eq:intermediateHSequation} and inserting $\lambda$, we have:
\begin{equation}
\label{eq:napprox}
e^{-\tau (H-\lambda)^2}=\frac{1}{\sqrt{\pi}}\sum_{j=-\frac{m}{2}}^{j=\frac{m}{2}} \sqrt{\frac{2}{m}}e^{-\frac{2j^2}{m}}e^{-2ij\sqrt{2 {\rm d}\tau}H} e^{-2ij\sqrt{2 {\rm d}\tau}\lambda}. 
\end{equation}
Following this, we note the orthogonality relationship for complex exponentials is given by
\begin{equation}
\sum_{\lambda =0}^{\lambda =K-1} e^{\frac{2i\pi\lambda}{K} (j-k) }= K\delta_{jk}.
\end{equation}
Next, for $\frac{m}{2}$ a square number such that  $K=\sqrt{\frac{m}{2}}$ is an integer, we have that $K\sqrt{2{\rm d}\tau}=\sqrt{\tau}$. Then, applying the complex exponential and summing over $\lambda$ on both sides, we have
\begin{equation}
\label{eq:DFT}
\sum^{\lambda = K-1}_{\lambda =0}e^{-\tau (H-\frac{\pi}{\sqrt{\tau}}\lambda)^2}e^{\frac{2i\pi\lambda k}{K}}=\frac{1}{\sqrt{\pi}}e^{-\frac{2k^2}{m}}e^{-2ik\sqrt{2{\rm d}\tau}H}.
\end{equation}
Defining $t=2k\sqrt{2{\rm d}\tau}$ and rearranging, we obtain an expression for the propagator:
\begin{equation}
\label{eq:ITQDEinvert}
    e^{-itH}=\sqrt{\pi}e^{\frac{t^2}{4\tau}}\sum^{\lambda = K-1}_{\lambda =0}e^{-\tau (H-\frac{\pi}{\sqrt{\tau}}\lambda)^2}e^{\frac{2 i\pi\lambda t}{\sqrt{\tau}}}.
\end{equation}

The fundamental motivation for such a representation is that it allows one to cast dynamical processes purely in terms of equilibrium information, with time acting only as a weighted phase on shifted Gaussian distributions. There are several contexts in which such an expression may prove useful, e.g., in ring-polymer molecular dynamics (RPMD) \cite{habershon_ring-polymer_2013}, a classical method that allows for calculating quantum imaginary time quantities via classical phase-space dynamics. A challenge associated with such techniques is that they are unable to capture the interference effects present in real-time quantum dynamics. It may, however, be possible to circumvent this by using RPMD to evaluate $e^{-\tau (H-\frac{\pi}{\sqrt{\tau}}\lambda)^2}$, and consequently characterise real-time quantum dynamics purely via a weighting of the classical phase-space calculation using Eq. \eqref{eq:ITQDEinvert}. 

The ITQDE formulation may also find applications in studies of complex quantum dynamics, such as quantum chaos \cite{PhysRevB.103.064309}. This phenomenon is often characterised by the spectral form factor \cite{PhysRevLett.128.190402}, which describes the fidelity between a coherent Gibbs state and its unitary time evolution \cite{PhysRevA.108.062201}. Such coherent Gibbs states may, however, be represented in real time by the ITQDE correspondence (or equivalently the propagator may be represented in imaginary time via Eq.\eqref{eq:ITQDEinvert}). This suggests the potential for such characterisations of chaos to be formulated in purely dynamical (or equilibrium) terms. Moreover, Eq.\eqref{eq:ITQDEinvert} may be useful in obtaining alternate formulations of dynamical quantities such as a system's dynamic response \cite{Kubo_1966} or  Out-Of-Time-Order Correlators (OTOCs)  \cite{Shenker2014}. This latter quantity is a measure of information scrambling, and recent proposals have demonstrated its intrinsic relationship to thermodynamics \cite{infothermo}. In this sense, Eq.\eqref{eq:ITQDEinvert} offers a natural route to further exploration of this connection.

Further, if we consider the continuous limit of Eq.~\eqref{eq:discretemaxmixedsoln}, another connection to open systems theory is revealed. Returning to Eq.~\eqref{eq:intermediateHSequation} and defining $\delta=\sqrt{\frac{2}{m}}$ and $x=j\delta$, the right-hand side of Eq.~\eqref{eq:intermediateHSequation} may be expressed as
\begin{equation}
\label{eq:Hs-discrete2}
e^{-\tau {H}^2}=\frac{1}{\sqrt{\pi}}\sum_{j=-\frac{m}{2}}^{j=\frac{m}{2}} \delta e^{-x^2}e^{-2ix \sqrt{\tau}{H}} +\mathcal{O}(\delta^2),
\end{equation}
such that in the limit $\delta \to 0$ we recover the \textit{Hubbard-Stratonovich} (HS) transformation \cite{PhysRevLett.3.77, ourpaper, ourpaper2,McCaul2021}:
\begin{align}\label{eq:hubbardstratonovich}
    e^{-\tau H^2} = \frac{1}{\sqrt{\pi}} \int_{-\infty}^{\infty} e^{-x^2} e^{-2ix\sqrt{\tau}H} {\rm d}x.
\end{align}

This result is usually understood as a correspondence linking a deterministic quadratic potential to a stochastic linear one. Here, however, it emerges from ITQDE as the continuous limit of infinitesimal dynamics, where the state space and its dual are oppositely evolved. Based on this limit, it is possible to infer a generalisation of the HS transformation by repeating the same limiting procedure, but beginning from a general initial state. Setting $\ket{\psi_{0}}=\sum_ja_j\ket{E_j}$, we find 
\begin{align}
&e^{-\tau(E_k+E_l)^2/4} \ket{E_k}\bra{E_l} \notag\\ &\quad =\frac{1}{2^n}\sum_{j=-\frac{n}{2}}^{j=\frac{n}{2}}{ n \choose \frac{n}{2}+{j}}e^{-2ij\sqrt{2 {\rm d}\tau}(E_k +E_l)}\ket{E_k}\bra{E_l}.
\end{align}
If we define the operator 
\begin{equation}
    \rho_{\rm mix}= \sum_{kl} a_ka_le^{-\tau(E_kE_l)/2} \ket{E_k}\bra{E_l},
\end{equation}
then the previous $\delta \to 0$ limit then yields
\begin{align}
  & e^{-\tau H^2/2}\rho_{\rm mix} e^{-\tau H^2/2} \notag\\
  &\quad = \frac{1}{\sqrt{\pi}} \int_{-\infty}^{\infty} e^{-x^2} e^{-ix\sqrt{\tau}H}\ket{\psi_0}\bra{\psi_0} e^{-ix\sqrt{\tau}H}  {\rm d}x.
\end{align}

Lastly, when it comes to further applications of ITQDE in the arena of quantum algorithms, the formulation described in Sec. \ref{sec:quantum computer} could be extended in the future to allow for estimating expectation values under thermal states associated with $H^2$, i.e., via an initial condition corresponding to the maximally mixed state per Eq. (\ref{eq:intermediateHSequation}). It may additionally be possible to consider thermal states under $H$, rather than $H^2$, if certain criteria on $H$ are satisfied, following Refs.  \cite{somma2013spectral,chowdhury_quantum_2016}. For example, Ref. \cite{chowdhury_quantum_2016} uses such a construction to design an LCU-based quantum algorithm for sampling from the thermal state $e^{-\beta H/2}$. This is enabled by a discrete Hubbard-Stratonovich transformation for approximating $e^{-\beta H/2}$ as a linear combination of unitaries, combined with a truncated Taylor series \cite{PhysRevLett.114.090502} to further approximate the constituent terms. This construction allows for implementing the approximation to $e^{-\beta H/2}$ coherently, i.e., as a single quantum circuit, which can subsequently be sampled from. We anticipate that an ITQDE-based algorithm developed for sampling from thermal states would share similarities to the algorithm in Ref. \cite{chowdhury_quantum_2016}, but would trade off the need for extra ancilla qubits and deep circuits for an increased sampling cost, i.e., where the number of circuits that are sampled from in an ITQDE-based framework would be higher, but the per-circuit implementation cost may be lower. Better understanding these tradeoffs, and their implications for the quantum computational resources required for running these algorithms at scale, would constitute valuable future work.

\section{\label{sec:outlook} Outlook} 
In this work, we have presented a method for constructing the solutions of non-unitary dynamics from unitary operations, which we have termed QDE. This was then applied in the context of imaginary time to derive the ITQDE correspondence between real and imaginary time evolution that lies at the heart of this work. Employing this, it was demonstrated that both ground states and spectra can be calculated from a single imaginary time trajectory, based on measurements of a set of dynamical overlaps.  The results presented here represent a first application of QDE; however, this technique is ripe for further refinement in terms of approximations to improve its efficiency, in the scope of its application, and in the understanding of its limitations. 

The performance of ITQDE depends upon the interplay of free parameters such as $\Delta\tau$ and $\lambda$. Looking ahead, it would be interesting to further probe these relationships, and to investigate how to improve the error scaling in ITQDE beyond what is obtained in Sec. \ref{sec:derivation}, e.g., via higher-order quadrature techniques, or by working backwards from the continuous limit of the HS transformation. In the latter case, the use of techniques such as Gaussian quadrature may be employed to obtain better discretisations, as sketched in App.~\ref{app:error_bounds}. Furthermore, while the majority of the calculations presented in this work use the large $m$ approximations in Eqs.~\eqref{eq:approx_shifted_expt} and \eqref{eq:approx_shifted_z} to avoid calculating very large factorials, the exponential weighting means that most of the weight is focused around the initial state and short-time overlaps. Embracing this feature may allow for truncated expressions with the potential to extrapolate out to much later $\tau$ than would otherwise be possible. 

We have additionally utilized ITQDE to develop a quantum algorithm for estimating Hamiltonian spectra. The algorithm operates by sampling a set of quantum state overlaps along a trajectory in imaginary time. It then uses the sampled data to construct an estimate of the ground state energy. In a classical post-processing step, the rest of the spectrum can then be obtained from the same set of data points sampled from a single trajectory in imaginary time. We have presented results of implementing this ITQDE-based quantum algorithm in IBM's superconducting qubit-based quantum hardware in order to estimate the spectrum of a two-qubit instance of the transverse-field Ising model. Looking forward, further implementation of error mitigation strategies and algorithmic techniques, e.g., \cite{stef1,stef2,KHANEJA200111}, may enable larger-scale spectral calculations via ITQDE on near-term quantum hardware. ITQDE-based quantum algorithms also represent an opportunity to impact other applications, such as topological data analysis \cite{stef4}. Looking further ahead, investigations of the asymptotic scaling of ITQDE-based algorithms targeted to implementation on fault-tolerant quantum computers would also constitute valuable future work. More broadly, the fact that ITQDE reconstructs spectra using only measurements of state overlaps means the method is, in principle, platform agnostic, and can be applied in any digital or analog setting where such measurements are feasible. 

We have also provided a preliminary sketch of potential uses for ITQDE in the context of stochastic and quantum thermodynamics. The presence of forward and reverse trajectories in ITQDE may allow for comparisons to be made to both classical and quantum fluctuation theorems, and we expect that continued study of these connections has the potential to yield new insights. 
In fact, while the work performed here has restricted itself to a quantum setting, Hilbert space representations of classical dynamics \cite{Chruscinski2006, PhysRevE.99.062121, McCaul_2022, Bondar2013a}, including the recently developed waveoperator representation \cite{waveoperator}, represent excellent prospects for deriving new correspondences between classical dynamical overlaps and distributions over the \textit{Koopman operator}. This is itself a vital tool in the analysis and control of highly nonlinear systems \cite{BEVANDA2021197}, and a classical equivalent of ITQDE---IT\textit{C}DE---might be developed to calculate such Koopman spectra directly from measurements. 

Beyond imaginary time, the overarching QDE methodology is expected to lend itself to a broader set of applications, limited only by the range of exotic states which might be expressible via an operator-valued differential equation, and the degree to which QDE renders them realisable and computable. A clear example of this could be the simulation of Lindbladian dynamics, both for the deliberate simulation of open system dynamics, e.g., on quantum computers, and for the mitigation of environment-induced errors \cite{PhysRevLett.131.110603}. We expect that further developing the framework for this and other applications will allow QDE to serve as a broadly useful tool in the study of non-unitary quantum dynamics into the future. 

\begin{acknowledgments}

We gratefully acknowledge discussions about this work with Christian Arenz and George Booth.  J.M.L. and A.B.M. are supported by Sandia National Laboratories’ Laboratory Directed Research and Development Program. This research used IBM Quantum resources of the Air Force Research Laboratory. G.M. is supported by the European Research Council (ERC) under the European Union’s Horizon 2020 Research and Innovation Program (grant agreement 833365), while D.I.B. is supported by the Army Research Office (grant W911NF-23-1-0288; program manager Dr.~James Joseph).
Sandia National Laboratories is a multimission laboratory managed and operated by National Technology \& Engineering Solutions of Sandia, LLC, a wholly owned subsidiary of Honeywell International Inc., for the U.S. Department of Energy’s National Nuclear Security Administration under contract DE-NA0003525. This article has been authored by an employee of National Technology \& Engineering Solutions of Sandia, LLC under Contract No. DE-NA0003525 with the U.S. Department of Energy (DOE). The employee owns all right, title and interest in and to the article and is solely responsible for its contents. The United States Government retains and the publisher, by accepting the article for publication, acknowledges that the United States Government retains a non-exclusive, paid-up, irrevocable, world-wide license to publish or reproduce the published form of this article or allow others to do so, for United States Government purposes. The DOE will provide public access to these results of federally sponsored research in accordance with the DOE Public Access Plan https://www.energy.gov/downloads/doe-public-access-plan. This paper describes objective technical results and analysis. Any subjective views or opinions that might be expressed in the paper do not necessarily represent the views of the U.S. Department of Energy, Army Research Office, or the United States Government. SAND2024-11565O.

\end{acknowledgments}

\bibliography{literature.bib}

\newpage 

\appendix
\section{Ground State Acceleration via $\lambda$}\label{app:accel}

\begin{figure}
    \centering
    \includegraphics[width=\linewidth]{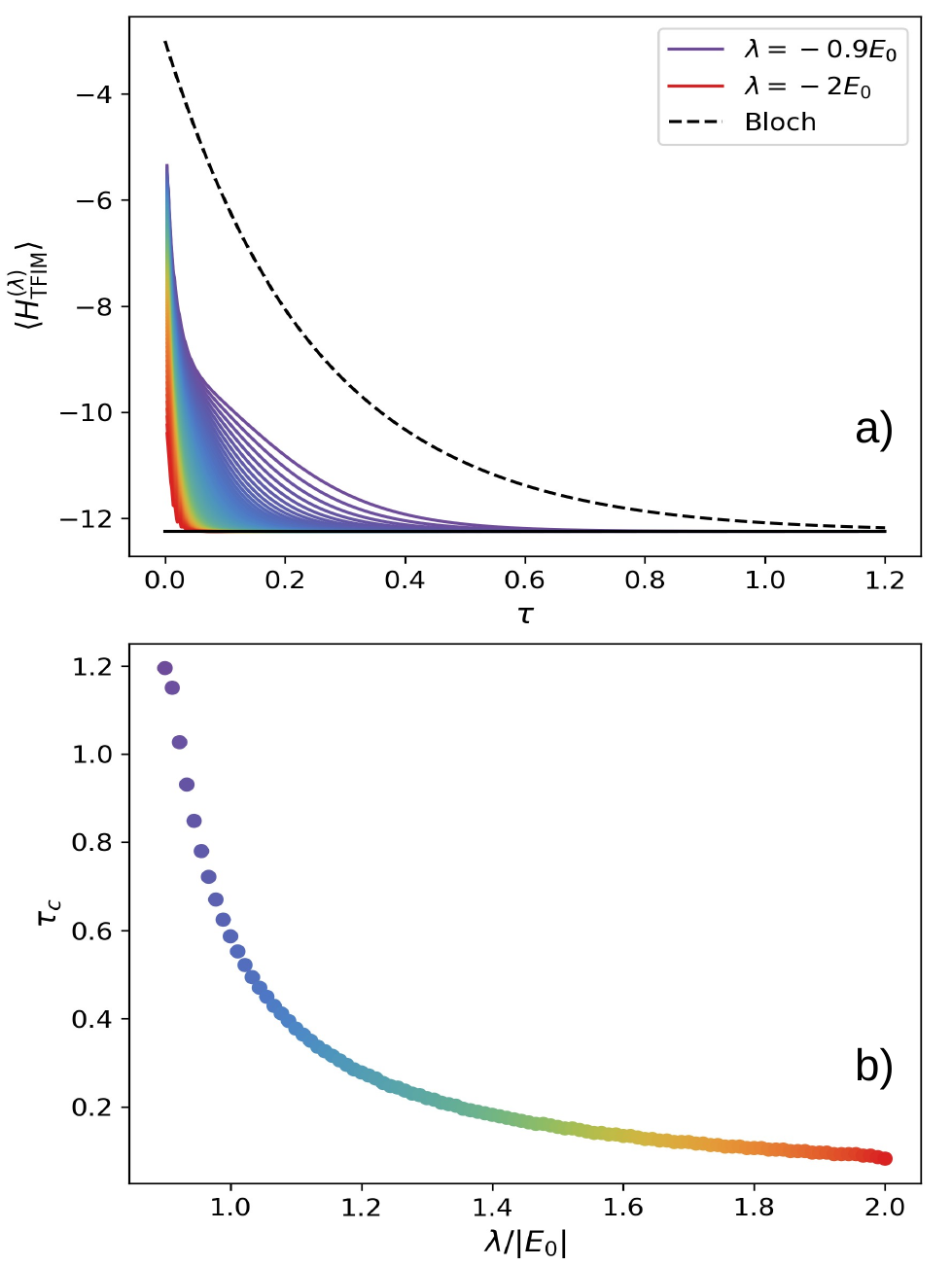}
    \caption{Illustration of the convergence to the ground state in a transverse field Ising model. In all cases, convergence is achieved significantly faster than evolution with the Bloch equation. The bottom panel demonstrates how $\tau_c$ - the temperature at which $|\langle{H}\rangle-E_0| <10^{-4}$ ($E_0$ being the ground state energy) - has the expected exponential dependence on $\lambda$. }
    \label{fig:convergence}
\end{figure}

Here, we discuss how tuning $\lambda$ can markedly improve the rate of convergence to the ground state with respect to $\tau$.  This phenomenon can be most easily illustrated with the case of a two level Hamiltonian.  We denote the energies of the two levels in terms of the average energy $\bar{E}$ and difference $2\Delta$, i.e. $E_0=\bar{E}-\Delta$ and $E_1=\bar{E}+\Delta$.  If one were to evolve via the Bloch equation i.e.
\begin{equation}
    \frac{d\rho}{d\tau} = -\frac{1}{2}\{\rho, H\}
\end{equation}
one would obtain (after normalisation) the Gibbs state
\begin{equation}
	\frac{1}{Z} e^{-\tau\left(H-\lambda\right)} = \frac{1}{Z}\left(e^{\tau\Delta}|E_0\rangle\langle E_0|+e^{-\tau\Delta}|E_1\rangle\langle E_1|\right).
\end{equation}
This implies that the ratio of probabilities for occupying the excited state $p(E_1)$ versus the ground state $p(E_0)$ is $\frac{p(E_1)}{p(E_0)} = e^{-2\tau\Delta}$ and that the shift parameter has no effect.  The value of $\tau$ to which one needs to evolve in order to obtain a good approximation to the ground state is determined by $\Delta$, with the ground state only being well-approximated when $\tau \sim \mathcal{O}(\frac{1}{\Delta})$.  For many complex systems, the low level energy states may be nearly degenerate and thus require an impractically large $\tau$ to obtain a good estimate for the ground state.  If one uses any $\lambda < E_0$, however, we find
\begin{align}
	\frac{1}{Z} e^{-\tau\left(H+\lambda\right)^2}  = \frac{1}{Z}&\bigg(e^{2\tau \Delta(\bar{E}+\lambda)}\ketbra{E_0}{E_0} \bigg. \notag \\ \bigg. +& e^{-2\tau \Delta(\bar{E}+\lambda)} \ketbra{E_1}{E_1}\bigg) \\ & \implies \frac{p(E_1)}{p(E_0)}=e^{-4\beta \Delta(\bar{E}+\lambda)}.
\end{align}
The rate of convergence to the ground state with respect to $\tau$ can therefore be tuned by $\lambda$, such that the required final value is reduced by a factor of $2\left(\bar{E}+\lambda\right)$.  This phenomenon is illustrated in Fig.~\ref{fig:convergence}, where decreasing $\lambda$ has the predicted exponential effect in the imaginary time $\tau_c$ at which $\rho^{(\lambda)}$ reaches the ground state, i.e. $|\langle H\rangle-E_0| < 10^{-4}$. 
\section{Additional Numerical Analyses} \label{app:diff_ic}

\begin{figure}
    \centering
    \includegraphics[width=\linewidth]{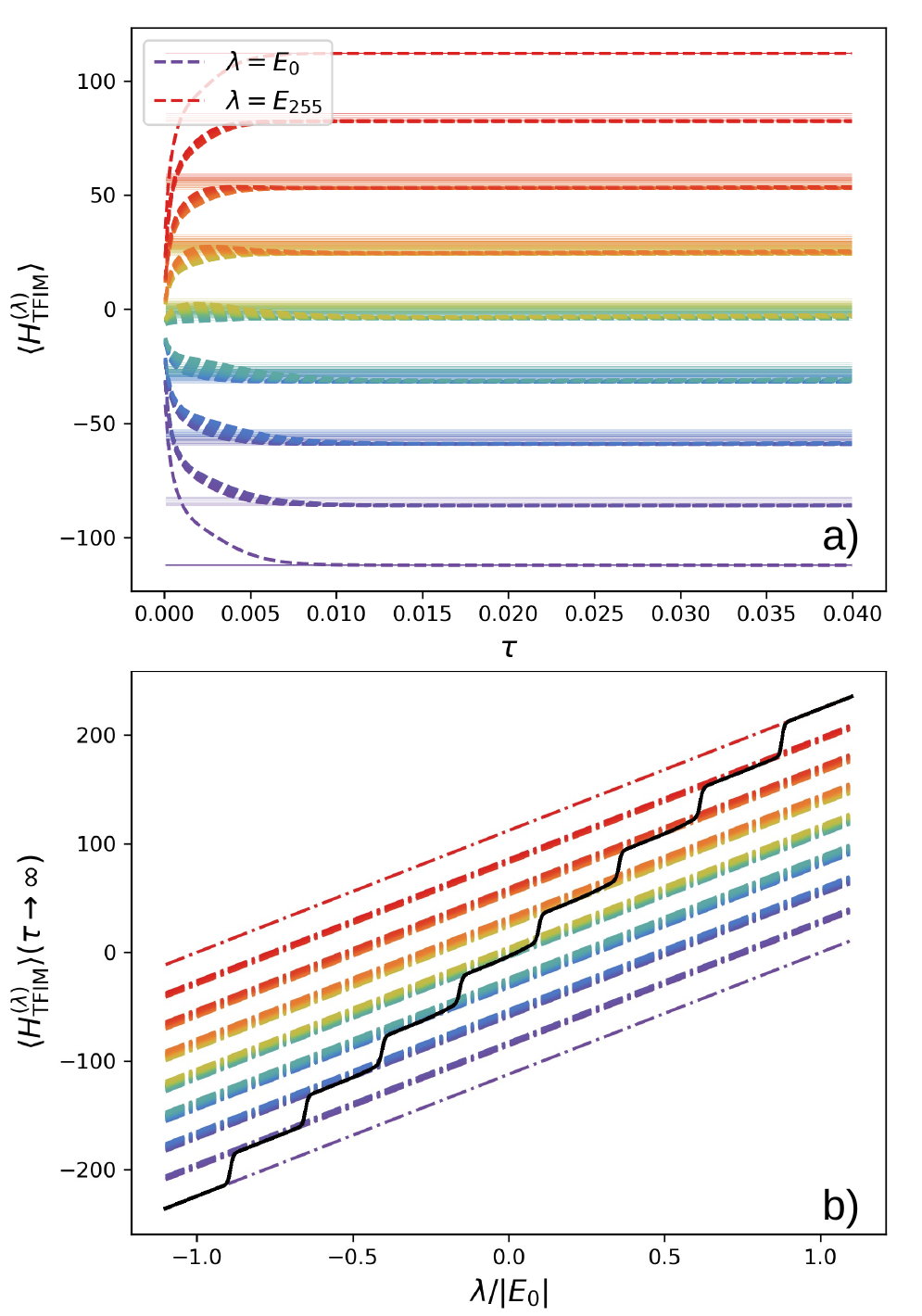}
    \caption{Results of spectral calculations performed for an eight site Ising Hamiltonian. Dashed lines indicate the eigenenergies of $H_{TFIM}$. Solid lines indicate the convergence of the spectral calculation to these eigenenergies. For a), we plot the calculated energy as $\lambda$ against propagation in imaginary time $\tau$ and find that our method converges to successively lower eigenstates of $H_{TFIM}$ before reaching the ground state $E_0$. For b), we plot the change in the steady state expectation value of $H_{TFIM}$ as $\lambda$ is varied, illustrating that a complete course-grained spectrum is obtainable using only expectation values estimated from a single trajectory only. }
    \label{fig:lambdasweepappendix}
\end{figure}

Here, we apply our method to compute the spectrum of a transverse-field Ising model and initialize the system in a single pure state $|\psi_0\rangle = \sum_{j}a_j|E_j\rangle$.  Results are plotted in Fig. \ref{fig:lambdasweepappendix}. 

In this case, we specifically initialize in the all-zero state given by $|\psi_0\rangle = |0\rangle^{\otimes 8}$ is used. This implies that $\rho^{(\lambda)}(\tau)$ has the form
\begin{equation}
	\rho^{(\lambda)}(\tau) = \sum_{ij}e^{-\frac{1}{4}\tau(E_i+E_j+2\lambda)^2}a_ia_j|E_i\rangle\langle E_j|.
\end{equation}

Consequently, when $\lambda$ is chosen such that $\langle H^{(\lambda)}\rangle(\tau\rightarrow\infty)$ converges to a band of closely spaced levels, it tends to converge towards the eigenstate corresponding to the lowest energy level in the band (i.e., rather than selecting out each level within the band separately). This occurs because of additional coherences in the energetic basis that are not present when $\rho^{(\lambda)}(0) \propto {1}$, thus reinforcing the notion that it is desirable to initialize the method accordingly in the maximally mixed state.

\section{Hadamard Test}\label{app:had_test}
Qubits are initialised in the state $\ket{0}$ before a Clifford gate is applied, in order to randomise the initial state such that over statistical sampling it mimics ${\rho}_S={1}$.  A Hadamard gate is also applied to the ancilla qubit, such that the overall state of the system at the first barrier is described by: 
\begin{align}
    \ket{\Psi_1} = \frac{1}{\sqrt{2}}(\ket{\psi_0}\ket{0} + \ket{\psi_0}\ket{1}),
\end{align}
where $\ket{\psi_0}$ denotes the randomised initial state of our system.  We then apply $2j$ copies of the unitary evolution operator ${\mathcal{U}}$ to $\ket{\psi_0}$ to simulate $2j$ steps of the evolution. Each of these operators is controlled by the ancilla qubit.  Then we have
\begin{align}
    \ket{\Psi_2} &= \frac{1}{\sqrt{2}}({\ket{\psi_0}\ket{0}} + {\mathcal{U}}^{2j}\ket{\psi_0}{\ket{1}}), \\
    &= \frac{1}{\sqrt{2}}({\ket{\psi_0}\ket{0}} + \ket{\psi_{2j}}{\ket{1}}).
\end{align}
In general, one needs to decompose an observable into a series of Pauli strings so that the expectation value can be evaluated efficiently.  Applying the Pauli string ${P}_\alpha$ controlled on the ancilla qubit yields
\begin{align}
    \ket{\Psi_3} &= \frac{1}{\sqrt{2}}({\ket{\psi_0}\ket{0}} + {P}_\alpha\ket{\psi_{2j}}{\ket{1}}).
\end{align}
Applying another $2j$ copies of ${\mathcal{U}}$ controlled on the ancilla then yields
\begin{align}
    \ket{\Psi_4} &= \frac{1}{\sqrt{2}}({\ket{\psi_0}\ket{0}} + \mathcal{U}^{2j}{P}_\alpha\ket{\psi_{2j}}{\ket{1}}).
\end{align}
The final Hadamard acting on the ancilla mixes the superposition again to give
\begin{align}
    \ket{\Psi_5} = &\frac{1}{2}\bigg(|0\rangle\big[|\psi_0\rangle + \mathcal{U}^{2j}P_\alpha|\psi_{2j}\rangle\big] \\ &+ |1\rangle\big[|\psi_0\rangle - \mathcal{U}^{2j}P_\alpha|\psi_{2j}\rangle\big]\bigg)
\end{align}
Now the probability of measuring $\ket{0}$ on the ancilla is
\begin{align}
    p_0 &= \frac{1}{4}\big(\langle\psi_0|+\langle\psi_{2j}|P_\alpha(\mathcal{U}^{2j})^\dag\big)\big(|\psi_0\rangle +\mathcal{U}^{2j}P_\alpha|\psi_{2j}\rangle\big), \\
    &= \frac{2 + 2\mathrm{Re}(\bra{\psi_{2j}}{P}_k\ket{\psi_{-2j}})}{4}.
\end{align}
From this, it is possible to obtain the real part of the overlap via a measurement of the ancilla, as
\begin{align}
    \mathrm{Re}(\bra{\psi_{2j}}{P}_k\ket{\psi_{-2j}}) = 2p_0 - 1.
\end{align}
Obtaining the imaginary part of this overlap can be achieved by modifying the circuit by applying a $S$ gate to the ancilla between the first Hadamard gate and $\mathcal{U}^{2j}$ gate.

\section{Estimator Method}\label{app:estimator}
\begin{figure}
    \centering
    \includegraphics[width=\linewidth]{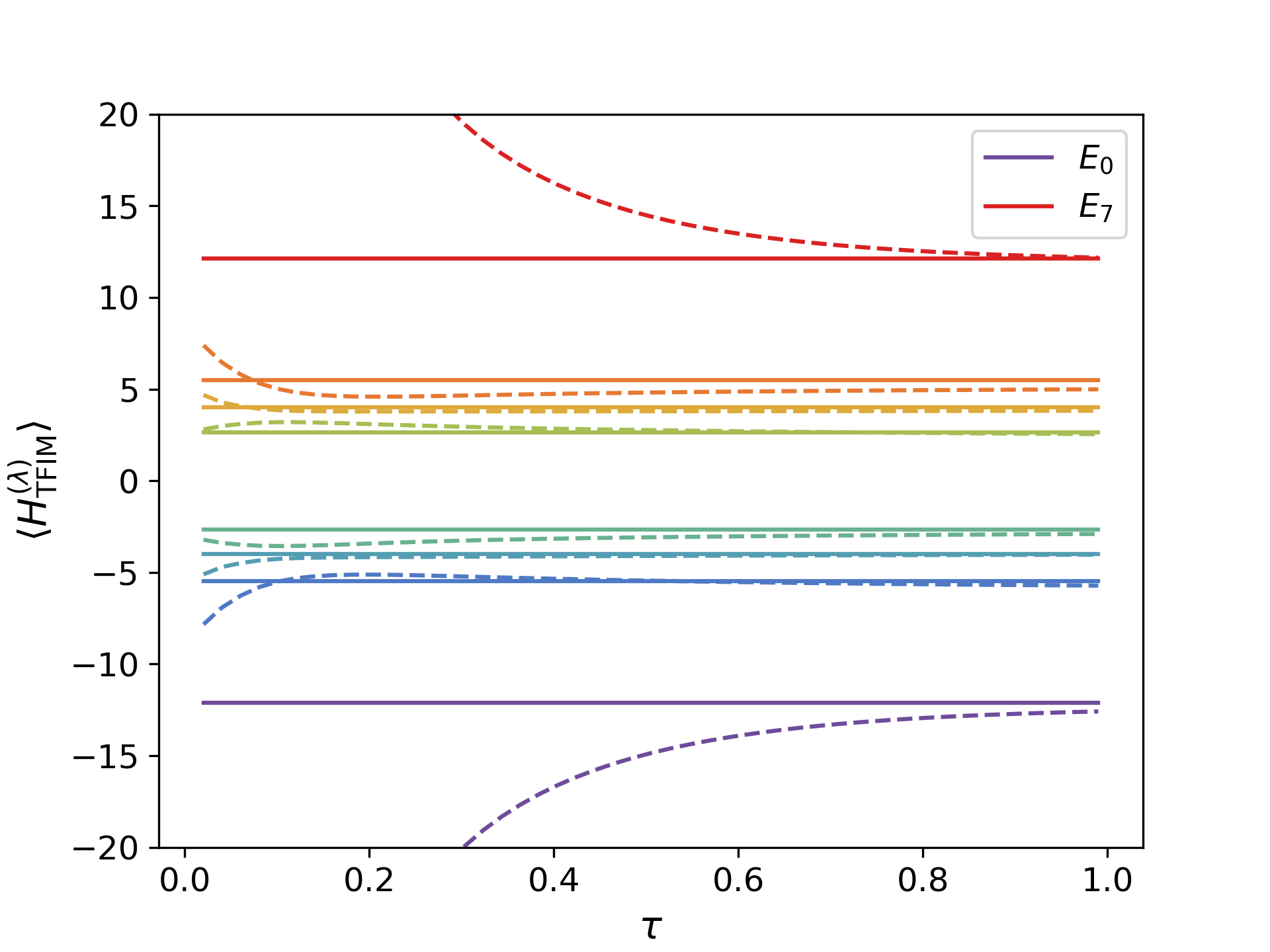}
    \caption{Energy expectation values for a 3 site transverse field Ising model with $J=1$, $h=2$ as the system propagates in imaginary time obtained.  The overlaps required for calculating Eqs.~\eqref{eq:approx_shifted_expt} and \eqref{eq:approx_shifted_z} are obtained using IBM's estimator primitive with the ibm qasm simulator to measure the expectation values of ${S}_{\mathrm{Re}}^j$ and ${S}_{\mathrm{Im}}^j$ as in Eq.~\eqref{eq:estimator_approach}}
    \label{fig:estimator_results}
\end{figure}
Another method of obtaining results for larger systems is to calculate the necessary overlaps directly from expectations.  Explicitly, the real and imaginary parts of the overlap $\langle\psi_{2j}|O|\psi_{-2j}\rangle$ can be expressed as expectations of the Hermitian operators $S_{\mathrm{Re}}^j$ and $S_{\mathrm{Im}}^j$ i.e.
\begin{align}
	S_{\mathrm{Re}}^j &= \frac{1}{2}\left({\mathcal{U}}^{2j} + ({\mathcal{U}}^{2j})^\dag \right), \\
	S_{\mathrm{Im}}^j &= \frac{1}{2i}\left({\mathcal{U}}^{2j} - ({\mathcal{U}}^{2j})^\dag \right),
\end{align}
such that
\begin{align}
	\mathrm{Re}/\mathrm{Im}\left(\langle\psi_{-2j}|O|\psi_{2j}\rangle\right) = \langle\psi_0|O{S}_{\mathrm{Re}/\mathrm{Im}}^j|\psi_0\rangle. \label{eq:estimator_approach}
\end{align}
This alternative approach reduces the number of qubits used in the simulation, which reduces the errors in the calculation due to noisy hardware.  However, this approach does not scale well in comparison to the Hadamard test approach discussed above.  This is because measuring the expectation values of ${S}_{\mathrm{Re}/\mathrm{Im}}^{j}$ in the circuit requires decomposing ${\mathcal{U}}^{2j}$ and $({\mathcal{U}}^{2j})^\dag$ into series of Pauli strings for every step.  With the Hadamard test approach, we only needed to perform the Pauli string decomposition $O$ once and then we could use it each time we evaluated the circuit.  We demonstrate the utility of this approach for a 1D transverse field Ising model with $J=1$, $h=2$, and $L=3$ by plotting the energy expectations against imaginary time in Fig.~\ref{fig:estimator_results}.  These results are obtained using IBM's qasm simulator with the estimator primitive backend rather than a real machine.  To mimic starting with the maximally mixed entangled state, we repeat the calculations with 50 different random initial states and average the calculated overlaps at each step.  The initial divergence in the trajectories of the energy expectation values highlight the possibility of encountering singularities when performing the calculations.  Though good convergence can often be obtained by sufficient evolution in $\tau$ despite these singularities, they can be avoided by reducing the step size in either $\tau$ or $\lambda$ such that the trajectories converge from below as seen in Figs.~\ref{fig:lambdasweep}, \ref{fig:FH_res}, \ref{fig:sampler_results}, \ref{fig:lambdasweepappendix}.  These results demonstrate that the alternative approach of measuring the expectations of $S_{\mathrm{Re}}^j$ and $S_{\mathrm{Im}}^j$ can be used to calculate the energy spectrum of the system for larger systems than the Hadamard test approach for near term quantum computers.  

\section{Error Bounds}\label{app:error_bounds}

One may obtain an error for ITQDE by considering the continuous limit of the correspondence, namely the Hubbard-Stratonovich transformation as obtained in Sec.~\ref{sec:stomp}.
Let us evaluate the integral via Gauss-Hermite quadrature (see Eqs. (3.5.15), (3.5.19), and (3.5.28) of Ref.~\cite{NIST:DLMF}),
\begin{align}
    & e^{-\tau H^2} \notag\\
    & = \frac{1}{\sqrt{\pi}} \sum_m \int_{-\infty}^{\infty} e^{-x^2} e^{-2i\sqrt{\tau}x E_m} \ket{E_m}\bra{E_m} dx \notag\\
    & = \frac{1}{\sqrt{\pi}} \sum_m \sum_{k=1}^n w_k e^{-2i\sqrt{\tau}x_k E_m} \ket{E_m}\bra{E_m} + {\mathcal{E}} \notag\\
    & = \frac{1}{\sqrt{\pi}} \sum_{k=1}^n w_k e^{-2i\sqrt{\tau}x_k H} + {\mathcal{E}},
\end{align}
where the coordinate grid $\{x_k\}$ and weights $\{ w_k \}$ are tabulated in Ref.~\cite{NIST:DLMF}. Moreover, there exists reals $\{ \xi_m \}$ such that the error term is
\begin{align}
    {\mathcal{E}} &= \frac{n! 2^n}{(2n)!} (-i\sqrt{\tau})^{2n} \sum_{m} E_m^{2n} e^{-2i\sqrt{\tau} \xi_m E_m} \ket{E_m}\bra{E_m} \notag\\
    &= \frac{n! 2^n}{(2n)!} (-i\sqrt{\tau})^{2n} \sum_{m, m'} E_{m'}^{2n} e^{-2i\sqrt{\tau} \xi_m E_m} \ket{E_m}\bra{E_m} \delta_{m, m'} \notag\\
    &= \frac{n! 2^n}{(2n)!} (-i\sqrt{\tau})^{2n} \sum_{m, m'} E_{m'}^{2n}\ket{E_{m'}}\bra{E_{m'}} e^{-2i\sqrt{\tau} \xi_m E_m} \ket{E_m}\bra{E_m} \notag\\
    &= \frac{n! 2^n}{(2n)!} (-i\sqrt{\tau})^{2n} H^{2n} \sum_{m}   e^{-2i\sqrt{\tau} \xi_m E_m} \ket{E_m}\bra{E_m}. \notag
\end{align}
To write it more compactly, we can say that there is a unitary operator ${\mathcal{U}}$ such that the error ${\mathcal{E}}$ is
\begin{align}
    {\mathcal{E}} = \frac{n! (2\tau)^n}{(2n)!} H^{2n}  {\mathcal{U}}.
\end{align}
The error is written in the polar operator form.

From Eq.~(5.6.1) of Ref.~\cite{NIST:DLMF}, we get
\begin{align}
    \sqrt{\frac{2\pi}{n+1}} \left(\frac{n + 1}{e}\right)^{n+1} < n!
    < \sqrt{\frac{2\pi}{n+1}} \left(\frac{n + 1}{e}\right)^{n+1} e^{\frac{1}{12n + 12}}.
\end{align}
Hence, for a unitary invariant matrix norm $\| \cdot \|$, we obtain the bound
\begin{align}
    \| {\mathcal{E}} \| < \frac{(n + 1)^{n + 1/2}}{(2n + 1)^{2n + 1/2}}e^{\frac{1}{12n + 12}} (2\tau e)^n \| H^{2n} \|,
\end{align}
which leads to
\begin{align}
    \| {\mathcal{E}} \| = \mathcal{O}\left( \left(\frac{\tau e}{2n}\right)^n \| H^{2n} \| \right), \quad n \to \infty.
\end{align}

\end{document}